\documentclass{aa}

\usepackage{graphicx}
\usepackage{txfonts}
\usepackage[colorlinks=true,allcolors=blue]{hyperref}

\usepackage{amsmath}    
\usepackage{amsmath,amssymb}
\usepackage{mathrsfs}
\usepackage[flushleft]{threeparttable}
\usepackage{siunitx}
\usepackage{xcolor}
\usepackage{booktabs, caption}
\usepackage{float}
\usepackage{pdflscape}
\usepackage{ulem}
\usepackage{multirow}
\usepackage{physics}
\usepackage{natbib}

\bibpunct{(}{)}{;}{a}{}{,}

\begin{document}

   \title{Kinematic detection of dusty outflows from active galactic nuclei: Polycyclic aromatic hydrocarbon kinematics of type 2 quasars with JWST/MIRI spectroscopy}

   \titlerunning{PAH kinematics of type 2 quasars}

   \author{Fergus R. Donnan\inst{1}\corrauth{fdonnan@ucsd.edu}
        \and Cristina Ramos Almeida\inst{2,3}\email{}
        \and Omaira Gonz\'alez Martín\inst{4}
        \and Karin Sandstrom\inst{1}
        \and Anelise Audibert\inst{2,3}
        \and Marina Bianchin\inst{2,3}
        \and Miguel Pereira-Santaella\inst{5}
        \and Ismael Garc\'ia-Bernete\inst{6}
        }

   \institute{$^{1}$ Department of Astrophysics, University of California San Diego, 9500 Gilman Drive, San Diego, CA 92093, USA\\
   $^{2}$ Instituto de Astrof\'isica de Canarias, Calle Vía Láctea, s/n, E-38205, La Laguna, Tenerife, Spain\\
   $^{3}$ Departamento de Astrof\'isica, Universidad de La Laguna, E-38206, La Laguna, Tenerife, Spain\\
   $^{4}$ Instituto de Radioastronomía y Astrofísica (IRyA), Universidad Nacional Autónoma de México, Morelia, Michoacán, Mexico\\
   $^{5}$ Instituto de F\'isica Fundamental, CSIC, Calle Serrano 123, 28006 Madrid, Spain\\
   $^{6}$ Centro de Astrobiolog\'{\i}a (CAB), CSIC--INTA, Camino Bajo del Castillo s/n, E--28692 Villanueva de la Ca\~nada, Madrid, Spain\\
   }

  \abstract
   {Active galactic nuclei (AGN) are thought to have dusty outflows, but unlike in the gas phase, it is challenging to measure the kinematics of dust. }
   {We present the detection and analysis of the kinematics of dust in five type 2 quasars at $z\sim0.1$ from the Quasar Feedback (QSOFEED) sample observed with JWST/MIRI spectroscopy.} 
   {We used principal component analysis tomography to produce velocity maps of polycyclic aromatic hydrocarbon (PAH) features, which are the smallest carbonaceous dust particles. We were then able to compare velocity maps of the PAHs with emission lines of ionised and molecular gas.}
   {We produced velocity maps of the 11.3 $\mu$m PAH feature, which traces large and neutral PAHs, for three out of the five objects, where all three show an outflow in the PAH kinematics. This becomes particularly clear after we subtracted disk kinematics, where the H$_2$ rotational transitions also show residuals consistent with an outflow. Compared to previous work with Seyfert galaxies, this work suggests that dusty outflows are more common at higher Eddington ratios, $\lambda_{\rm Edd}\gtrsim0.1$, in agreement with previous suggestions, although the sample size is small. We were unable to produce velocity maps for the 6.2 $\mu$m PAH, which traces ionised PAHs, potentially due to differences in the intrinsic profile and/or suppression of the feature in AGN, which was seen previously in Seyfert galaxies. This reflects studies of PAH band ratios where AGN outflows have more neutral PAHs. }
   {This work demonstrates that dusty outflows may be common, particularly at high Eddington ratios, and it therefore plays a key role in the evolution and life cycle of AGN.}


   \maketitle
    \nolinenumbers

\section{Introduction}

Active galactic nuclei (AGN) are known affect their host galaxy through feedback processes, and they thus play a pivotal role in the evolution of galaxies \citep[e.g.][]{Kormendy2013, Heckman2014}. 
Because the majority of AGN are obscured by dust \citep[e.g.][]{RamosAlmeida2017,Hickox2018, Boorman2025}, many key questions exist about the nature and lifetime of dust around AGN. A typical picture of galaxy evolution suggests that as AGN activity is triggered, such as through galaxy galaxy mergers, a dust-obscured phase of rapid supermassive black hole growth occurs \citep[e.g.][]{Sanders1988, Hopkins2008, Ricci2022}. The subsequent radiation field of the accretion disk is then able to clear the obscuring dust, resulting in dusty outflows \citep[e.g.][]{Fabian2008, Ishibashi2018, Arakawa2022} and leading to AGN that can be observed in the optical. Moreover, the relatively recent discovery of numerous little red dots (e.g. \citealt{Hviding2025, Kocevski2025, Rusakov2026, PerezGonzalez2026}), which show high column density gas obscuring a supermassive  black hole, but little dust \citep[][]{Barro2025, Casey2024, Chen2025, Delvecchio2025} despite the presence of metal emission lines \citep[e.g.][]{Matthee2024}, adds to this story, further suggesting that AGN may clear their dust \citep[e.g.][]{ Matthee2026}. Therefore, the expulsion of dust through AGN-driven winds is a key process for understanding how AGN evolve and inject material into the interstellar medium.

Extended dust in the polar axis of AGN were detected via mid-infrared imaging \citep[][]{Braatz1993, Bock2000, Radomski2002, Radomski2008, Packham2005, Reunanen2010, Asmus2014, Asmus2016, Garcia-Bernete2016, Asmus2019, Stalevski2017, Haidar2024, Haidar2026, Campbell2025} and with infrared interferometry in particular, providing the high angular resolution needed to image AGN tori, revealing polar structure perpendicular to the tori  \citep[][]{Raban2009,Honig2012, Honig2013, Lopez-Gonzaga2014, Tristram2014, Lopez-Gonzaga2016,  Leftley2018, Isbell2022, GamezRosas2022}. The first kinematic evidence of dusty outflows from AGN was presented in \citet{Donnan2026} through the measurement of the velocity of polycyclic aromatic hydrocarbon (PAH) features in the infrared.

PAHs are the smallest carbonaceous dust grains, consisting of aromatic rings of carbon with hydrogen bonds, and they make up $\sim 1\%-5\%$ of the mass of dust in galaxies \citep[e.g.][]{Draine2007} but can contribute up to $20\%$ of the luminosity of dust grains \citep[e.g.][]{Smith2007} in metal-rich galaxies. After excitation from a UV photon, these molecules relax, where the C-C and C-H bonds bend and stretch at various frequencies, giving rise to broad emission features in the infrared (3-17 $\mu$m; e.g. \citealt{Tielens2008, Li2020}). Unlike the thermal continuum from stochastically heated dust grains, which dominate the continuum emission in the infrared, it is possible to measure the Doppler velocity shifts of PAH features, although there are significant challenges. Typically, PAHs are fitted using an integrated spectrum such as \textsc{PAHFIT} \citep{Smith2007}, \textsc{CAFE} \citep{Marshall2007, Diaz-Santos2025}, or \textsc{SPIRIT} \citep{Donnan2024} due to their broad profiles. 

PAHs are typically observed to have a lower equivalent width in AGN \citep[e.g.][]{Armus2007, Spoon2007, Alonso-Herrero2014, Ramos2014, Donnelly2024, Lai2023, RamosAlmeida2023, Garcia-Bernete2024b} because of the strong dust continuum from the torus and/or destruction of the PAHs through processes such as photo-destruction or shocks \citep[e.g.][]{Siebenmorgen2004, Garcia-Bernete2022c, Garcia-Bernete2024b, Zhang2023, Zhang2024, Zhang2026}. The PAH band ratios in AGN outflows show more neutral and larger PAHs, consistent with the idea of preferential destruction of small and ionised PAHs \citep[e.g.][]{Garcia-Bernete2022c, Garcia-Bernete2024b}.

There are two main challenges to measuring PAH kinematics. First, the features are significantly broader than any velocity shift, and second, the shape of the intrinsic feature profile is not well defined and can vary depending on the excitation conditions and properties of the PAH molecules \citep[e.g.][]{Peeters2002, Candian2015, Shannon2019, Canelo2026}. For these two reasons, traditional methods of measuring kinematics such as modelling each spaxel and fitting for a velocity are not feasible. Instead, \citet{Donnan24b} showed how principal component analysis (PCA) tomography \citep[][]{Steiner2009} can be used to measure PAH kinematics. They presented the first velocity maps of PAH features. This work was followed by \citet{Donnan2026}, where the first kinematic detection of outflowing dust from AGN was presented.

The sample studied in \citet{Donnan2026} was restricted to Seyfert galaxies with fairly low Eddington ratios ($\log\lambda_{\rm Edd} \lesssim -1.5$) and AGN luminosities ($\log L_{\rm bol} \lesssim
 44$ erg s$^{-1}$). In this work, we apply the same techniques to a sample of five optically selected type 2 quasars (QSO2s) from the Quasar Feedback (\href{https://research.iac.es/galeria/cristina.ramos.almeida/qsofeed/}{QSOFEED}) sample \citep{RamosAlmeida2022,Pierce23,Bessiere24}. These QSO2s occupy a parameter space of higher AGN luminosity and higher Eddington ratio ($\log L_{\rm bol} \gtrsim
 45.5$ erg s$^{-1}$ and $\log\lambda_{\rm Edd} \gtrsim -1.0$; \citealt{RamosAlmeida2025, Ramos2026}). Unlike the majority of Seyfert galaxies, which are found in the permitted region of the column density-Eddington ratio diagram \citep{Fabian2008, Ricci2017b}, the five QSO2s are in either the blowout or polar dusty winds regions, indicating that they must be actively clearing gas and dust from their nuclear regions \citep{Ramos2026}. Incidentally, the \cite{Ramos2026} successfully reproduced the JWST MIRI nuclear spectrum of one of them with the CAT-3D wind model of \citet{Honig2017}, which models a clumpy torus with a dusty wind component. Therefore, these targets are ideal candidates for studying their PAH kinematics and comparing them with the results found for lower-luminosity AGN \citep{Donnan2026}.

The paper is structured as follows. In Sect. \ref{sec:Data} we describe the QSOFEED sample and the MIRI MRS observations with JWST. In Sect. \ref{sec:Methods} we describe the method of PCA tomography and kinematic modelling of the disk kinematics. In Sect. \ref{sec:Results} we present velocity maps for each target. Finally, in Sect. \ref{sec:Discussion} we discuss our results. Throughout this work, we assume a $\Lambda$ cold dark matter cosmology with $H_0 = 70$ km s$^{-1}$ Mpc$^{-1}$, $\Omega_m = 0.27$, and $\Omega_{\Lambda} = 0.73$.

\section{Sample, observations, and data reduction}
\label{sec:Data}

We used MIRI MRS observations from the JWST general observer program 3655 (PI: Ramos Almeida; MAST \href{https://archive.stsci.edu/doi/resolve/resolve.html?doi=10.17909/8w9h-re72}{{doi:10.17909/8w9h-re72}}). Five optically selected QSO2s with redshifts 0.09 $\leq$ z $\leq$ 0.12, AGN luminosities of log L$_\text{bol} = 45.5-46.0$ erg~s$^{-1}$, and stellar masses of log M$_*$ = 10.9-{11.3} M$_\odot$ were observed as part of this programme (see Table \ref{tab:Sample}). They are representative of the gas-rich QSO2s in the QSOFEED sample, with total molecular gas masses of M$_{\rm H_2}$=4-18$\times$10$^{9}$ M$_{\sun}$ and evidence of cold molecular outflows detected from Atacama Large Millimeter Array (ALMA) CO observations \citep{RamosAlmeida2022,Audibert25}. They also have some of the most extreme ionised gas kinematics within the QSOFEED sample \citep{Bessiere24,Speranza24}. 

The mid-infrared nuclear spectra of the five QSO2s, obtained from the JWST MIRI observations used here, were first presented in \citet{RamosAlmeida2025,Ramos2026}. They revealed different continuum shapes and silicate feature strengths, including emission features, even though they are obscured quasars. Their [NeV]/[NeII] ratios range from 0.1 to 2.1 and [NeIII]/[NeII] from 1.0 to 3.5, indicating different coronal line and ionising continuum strengths. The equivalent widths of the PAH bands range from less than 0.002 to 0.075 $\mu$m, from which \citet{RamosAlmeida2025} measured nuclear star formation rates of $\le$3-7 M$_{\sun}$~yr$^{-1}$ and a higher fraction of neutral PAHs over the ionised ones.

The MIRI MRS observations were made from May 2024 to January 2025 using a four- and two-point dither sequence for the target and background observations, respectively (we refer to \citet{RamosAlmeida2025} for details on the observations). The data reduction was made as described in \citet{Gonzalez2025}, using version 1.14.1 of the JWST pipeline \citep{Bushouse2024}. After the data cubes were reduced, we subtracted the unresolved point spread function (PSF) from them using the tool {\it MRSPSFisol} \citep{Gonzalez2025}. This step is key to studying the underlying extended emission of the QSO2s, as their high luminosities can result in a very strong point source from the hot dust continuum of the torus. The analysis presented in the following sections was conducted using the PSF-subtracted cubes, but all the velocity maps presented here were also created with the original non-PSF-subtracted ones for comparison. The original cubes produce much noisier maps, but are generally similar. This is likely because the PCA detects small wiggles and artefacts as bright features because the PSF is relatively bright compared to the circumnuclear emission. For the PSF-subtracted cubes, the PCA detects any artefacts well, resulting in a clean velocity component (see Sect. \ref{sec:Methods}). 

\begin{table*}                    
  \caption{Objects from the QSOFEED sample observed with JWST/MIRI MRS.}
  \label{tab:Sample}
    \def\arraystretch{1.2}
    \setlength{\tabcolsep}{2pt}

    \begin{threeparttable}
\makebox[\textwidth][c]{%
  \begin{tabular}{cccccccccccc}

    \hline

    SDSS ID & Short ID & RA & Dec & $z$ & $\log L_{\rm bol}$ & $\log\lambda_{\rm Edd} $ & $\log M_{\rm BH}$ & $\log N_{H}$& $i$ &  PA &  \\
    & & & & & erg s$^{-1}$ & & $M_{\odot}$ & cm$^{-2}$ & $^{\circ}$ & $^{\circ}$ \\
    (1) & (2) & (3) & (4) & (5) &  (6) & (7) & (8) & (9) & (10) & (11)\\

    \hline
J101043.36+061201.4 & J1010 & 10:10:43.36 & +06:12:01.4 & 0.0977 & 45.6 & $-0.8 \pm 0.8$ & $8.4 \pm 0.8$  & 22.3 & $37\pm1$ & $287 \pm8$ \\
J110012.39+084616.3 & J1100 & 11:00:12.39 & +08:46:16.3 & 0.1004 & 45.9 & $0.0 \pm 0.5$ & $7.8 \pm 0.4$ & 22.6 & $38\pm1$ & $68\pm3$   \\
J135646.10+102609.0 & J1356 & 13:56:46.10 & +10:26:09.0 & 0.1232 & 45.5 & $-1.0 \pm 0.4$ & $8.6 \pm 0.3$ & 23.0 & $54\pm1$ & $107\pm11$  \\
J143029.88+133912.0 & J1430 & 14:30:29.88 & +13:39:12.0 & 0.0851 & 45.8 & $-0.4 \pm 0.4$ & $8.2 \pm 0.4$ & 22.9& $38 \pm 2$  &  $27 \pm3$  \\
J150904.22+043441.8 & J1509 & 15:09:04.22 & +04:34:41.8 & 0.1115 & 46.0 & $-0.2 \pm 0.8$ & $8.3 \pm 0.8$ & 23.1 & $43\pm1$  &  $266 \pm 4$ \\

    \hline
  
  \end{tabular}
  }
\tablefoot{(1): SDSS source ID.  (2): Short ID. (3): Right ascension. (4): Declination. (5): Redshift. (6): AGN bolometric luminosity from extinction-corrected [O III] luminosities, from \citet{Ramos2026}. (7): Eddington ratio from \citet{Kong18}. (8): Black hole mass from \citet{Kong18}. (9): Column density from CO \citep{Ramos2026}.
(10): Disk inclination. (11): Disk position angle. The disk inclination and position angles correspond to those derived from modelling of CO(2-1) data from \citet{Audibert25}, except in the case of J1430 (PA=$3 \pm1$ in \citealt{Audibert25}), where the position angle was inferred from the [NeII] velocity map presented here.}
  \end{threeparttable}
 \end{table*}

\section{Methods}
\label{sec:Methods}
\subsection{PCA tomography}
The application of PCA to astronomical integral field unit data cubes was first presented by \citet{Steiner2009}, who used PCA tomography to linearly decompose the data cube into orthogonal components. These components consist of specific spectral features that have some spatial distribution, ordered such that the first principal component contains the greatest variance in the data, with subsequent components containing progressively less variance. Typically, higher-order components contain only noise.

\begin{figure}
        \includegraphics[width=\columnwidth]{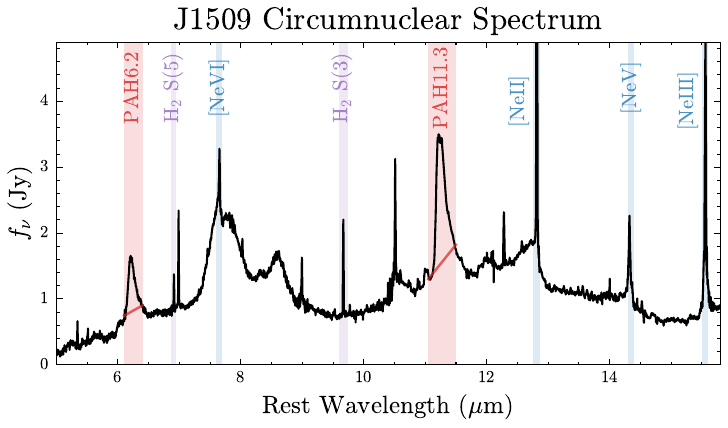}
    \caption{Circumnuclear spectrum of J1509, extracted through a 2'' radius aperture from the PSF-subtracted cubes. We highlight the emission features we analysed. The local continua used for the PAH features are also shown. }
    \label{fig:Spec} 
\end{figure}

PCA tomography was first shown to be able to measure the kinematics of PAH features by \citet{Donnan24b}, where the first principal component typically contains the rest-frame emission and the second component contains the velocity information. Before applying the PCA decomposition to the data cube of a given emission feature, we first subtracted a local continuum around the feature. This helped us to isolate the emission feature as the continuum can have a different spatial distribution to the emission feature and can thus confuse the PCA decomposition, where it instead identifies regions of continuum versus emission feature rather than kinematics. We constructed a local continuum by masking the emission feature and fitting a straight line in $f_{\nu}$ to the 15 wavelength channels of either side of the sub-cube. This is an approximate continuum and did remove some of the PAH flux in the wings. However, as we are interested in the kinematics and not the fluxes of the PAH features, this simple continuum is sufficient. We explore the continuum subtraction further in Appendix \ref{sec:ContSub}. The continua used for the 6.2 $\mu$m and 11.3 $\mu$m PAH features are shown in Fig. \ref{fig:Spec}. 

After continuum subtraction, the subtracted cube had the form $\mathbf{I}_{x, y, \lambda}$. We applied smoothing using a Gaussian kernel with $\sigma=1$ pixel and then masked pixels with a low signal-to-noise ratio following the process described in \citet{Donnan2026}. The cube was then collapsed into two dimensions, where the spatial dimensions, $x, y$ were transformed by
\begin{equation}
\label{eqn:CoordTrans}
    \beta = \mu (x-1) +y,
\end{equation}
where there are $\mu\times\nu$ total spatial pixels, 
resulting in the data having the form of a 2D array,  $\mathbf{I}_{\beta, \lambda}$. The PCA was applied to this array, where 
\begin{equation}
    \mathbf{T}_{\beta, k} = \mathbf{I}_{\beta, \lambda} \cdot \mathbf{E}_{\lambda, k}, 
\end{equation}
where $\mathbf{T}_{\beta, k}$ are the tomograms of each principal component, $k$, which is the spatial distribution of a given spectral feature, $\mathbf{E}_{\lambda, k}$, which are known as eigenspectra.

We constructed a velocity map from the PCA decomposition by taking the eigenspectrum of the first principal component as the rest-frame emission profile and measuring the wavelength shift and thus velocity required to reproduce the spectral profile of each spaxel on the reconstructed data cube using the first four principal components (higher-order components are dominated by noise, containing $\lesssim 1\%$ of the variance; \citealt{Donnan24b}). More details of this process can be found in \citet{Donnan2026}. To produce velocity error maps, we repeated the process 50 times, resampling the data each time from the error bars of the data cube, and took the standard deviation of the 50 samples to obtain an error estimate of the velocity.

\begin{figure}
        \includegraphics[width=\columnwidth]{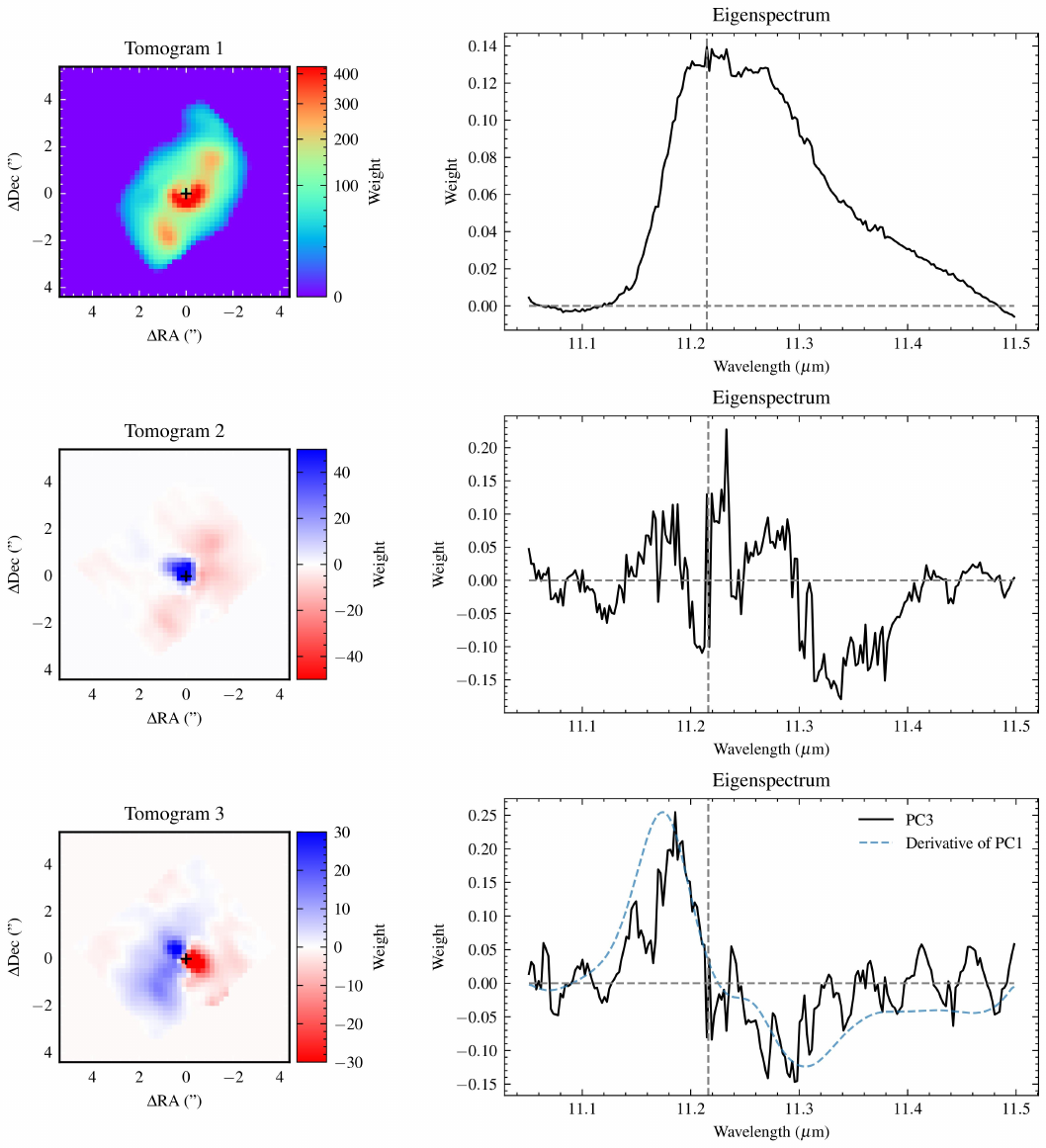}
    \caption{PCA decomposition of the 11.3 PAH feature for J1509. \textit{Right panels}: Eigenspectra of each component. \textit{Left panels}: Tomograms that describe how the eigenspectra map onto the spatial axes. The dashed blue line shows the derivative of the eigenspectrum of the first principal component, which is expected to match the eigenspectrum of the component containing velocity information, in this case, component 3. The second component detects artefacts in the nucleus due to the PSF subtraction. The vertical dashed line shows where the third eigenspectrum crosses from positive to negative.}
    \label{fig:PCA} 
\end{figure}

We demonstrate the PCA decomposition as applied to the 11.3 $\mu$m PAH for J1509 in Fig. \ref{fig:PCA}. The velocity information is contained in the third principal component, where the second detects artefacts from the PSF subtraction. We verified that we traced the velocity by inspecting the shape of the third principal component, which should match the derivative of the first component in the case of a Doppler shift. This is because PCA is a linear decomposition, and we can write a Doppler shift as a linear function after a Taylor expansion,
\begin{equation}
     f_{\rm obs}\left(\lambda\right) \approx f_{\rm rest}\left(\lambda\right) + \frac{v}{c}\frac{df_{\rm rest}}{d\lambda} \lambda  \approx a_i^{(1)}\phi_1(\lambda) + a_i^{(2)}\phi_2(\lambda) +  .., 
\end{equation}
where $f_{\rm obs}$ is the observed emission feature at velocity, $v$, where $f_{\rm rest}$ is the rest-frame feature. The left-hand side is the PCA decomposition, where $a_i^{(j)}$ is the tomogram of component $j$, and $\phi_j(\lambda)$ is the corresponding eigenspectrum. This shows that we expect the first component to be rest-frame emission and a higher-order component to contain the velocity information with an eigenspectrum proportional to the derivative of the rest frame profile. This is indeed what we observe in Fig. \ref{fig:PCA}, where the PSF artefacts add into the data linearly (as PSF subtraction is a linear process, this makes sense) in addition to the Doppler shift. The artefacts appear as component 2 rather than 3 because they contain more variance in the original data than the velocity (this is shown in Fig. \ref{fig:PCA}, where tomogram 2 had higher weights than tomogram 3).

We applied the PCA decomposition and constructed velocity maps for ionised and molecular gas emission lines as well as the 6.2 $\mu$m and 11.3 $\mu$m PAH features as these features are sufficiently bright and narrow enough. For the ionised gas, we used the [NeII] (12.81 $\mu$m, 41 eV), [NeV] (14.32 $\mu$m, 126 eV), and [NeVI] (7.65 $\mu$m, 158 eV) lines to trace the low and highly ionised gas respectively. For the molecular gas, we used the H$_2$ S(1) (17.03 $\mu$m) and the H$_2$ S(3) (9.66 $\mu$m) rotational transitions of molecular hydrogen. These emission features are highlighted in Fig. \ref{fig:Spec}.

\subsection{Disk modelling}
\label{sec:DiskModelling}
The position angle (PA) and inclination of the circumnuclear disk was well constrained for each of the five quasars by fitting a disk model to CO observations \citep[][]{RamosAlmeida2022, Audibert25}. This allowed us to model and subtract the disk kinematics to reveal any non-circular motions due to inflowing or outflowing material \citep[e.g.][]{Davies2024, Donnan2026}.

We followed the disk modelling presented in \citet{Veenema2025} and \citet{Donnan2026}, where we fitted the 2D velocity map of each emission feature with a fixed inclination, $i$, and position angle, $\phi$, while fitting for a radial dependent rotation curve, $V(R)$, parametrised by a $\tanh$ function that allows the rotation curve to rise steeply where $R<R_{\textrm{turn}}$ and flatten at $R>R_{\textrm{turn}}$. The rotation curve has the form
\begin{equation}
    V(R) = V_{\textrm{max}}\tanh{\left(\frac{R}{R_{\textrm{turn}}}\right)},
\end{equation}
where the $V_{\textrm{max}}$ and $R_{\textrm{turn}}$ are free parameters. The observed velocity, $V_{\rm obs}(x,y)$, of each spaxel $(x,y)$ is therefore 
\begin{equation}
    V_{\rm obs}(x,y) = V_{\rm sys} + V(R)\,\sin i \,\cos\theta,
\end{equation}
where $V(R)$ is the velocity of each ring at radius $
R$, and $V_{\rm sys}$ is the systemic velocity. The radius, $R$, is the intrinsic deprojected radius that relates to the coordinates of each spaxel $(x,y)$ via
\[
\begin{aligned}
x' &= x\cos\phi + y\sin\phi, \\
y' &= -x\sin\phi + y\cos\phi, \\
R &= \sqrt{x'^2 + \left(\frac{y'}{\cos i}\right)^{2}}, \\
\cos\theta &= \frac{x'}{R}. \\
\end{aligned}
\]

We fitted the 2D disk model to the velocity map using Markov chain Monte Carlo sampling from \textsc{NUMPYRO} \citep{Phan2019}. We fixed the inclination and position angle to the values shown in Table \ref{tab:Sample}, which were measured from CO data in \citet{Audibert25}. We fixed the orientation of the disk as it is difficult to constrain from a velocity map alone if there are multiple components such as outflows and bars. We did this for each galaxy, but as we discuss in Sect. \ref{sec:Results}, only three of the five targets show a reliable detection of kinematics in the PAH features, namely J1100, J1430, and J1509, likely due to the low signal-to-noise ratio in the two remaining targets.

\section{Results}
\label{sec:Results}

\subsection{Velocity maps}
\label{sec:velocity_maps}
\begin{figure*}
        \includegraphics[width=\textwidth]{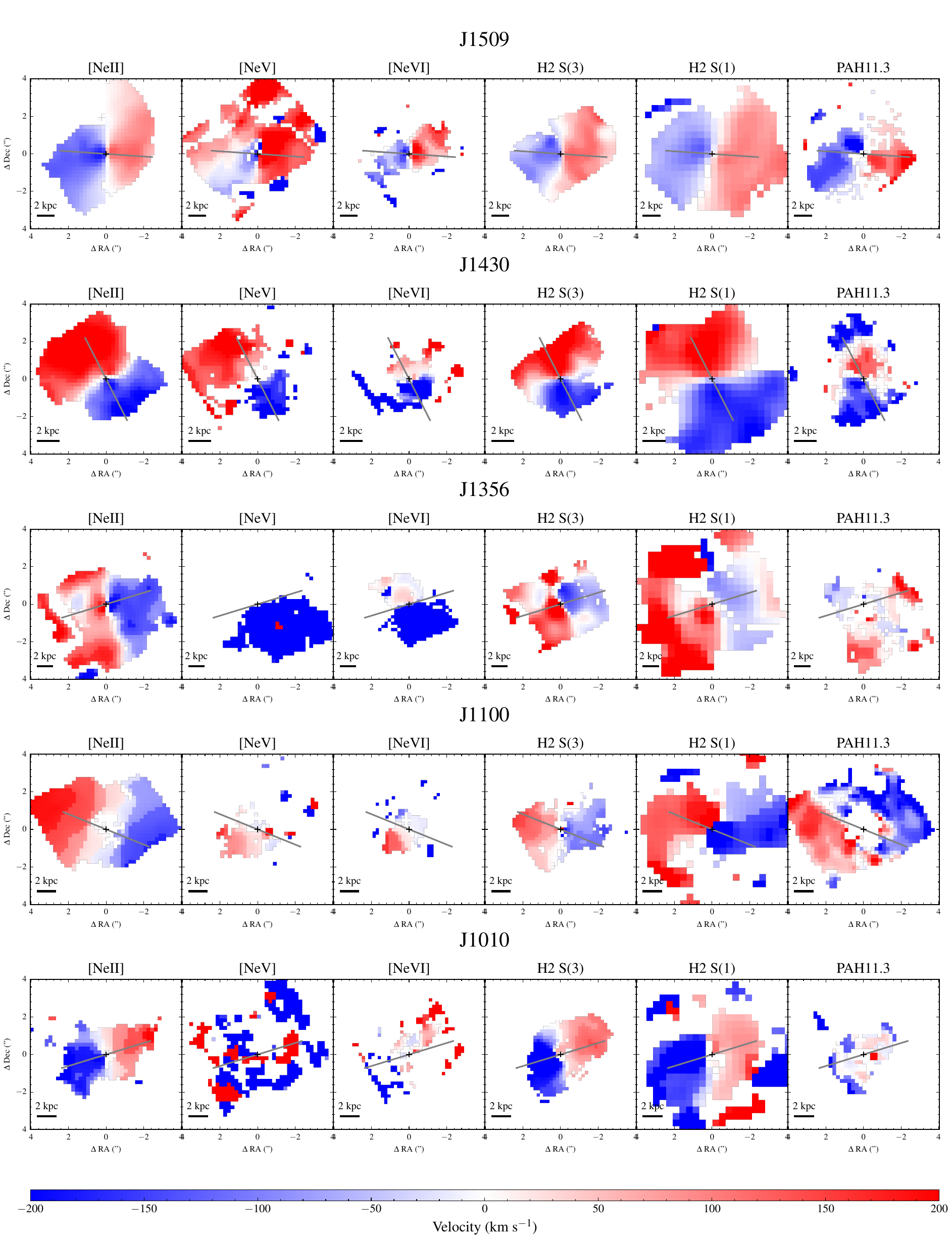}
    \caption{Velocity maps of each spectral feature for each galaxy derived from the PCA decomposition. We only show spaxels where the velocity was measured to $> 1\sigma$. Each panel has the same velocity scale from -200 - 200 km s$^{-1}$. The solid grey line shows the direction of the major axis of the galaxy disk for each galaxy and can be found in Table \ref{tab:Sample}.}
    \label{fig:VelMaps} 
\end{figure*}

We present velocity maps for each for the five QSO2s in Fig. \ref{fig:VelMaps} of ionised gas, molecular gas, and the 11.3 $\mu$m PAH. We also show the position angle of the large-scale galaxy disk as inferred from CO data \citep[][]{RamosAlmeida2022, Audibert25}. We find clear differences in the position angle of the kinematic axis for different features, likely reflecting different contributions of outflowing, inflowing and rotating material. In the following sections, we discuss the kinematics of the ionised, molecular, and dust phases for the five objects.

\subsubsection{Ionised gas}
The kinematics of the low-ionisation potential [NeII] line trace the gas disk well, matching the position angle of the CO data, except in the case of J1430 (see Table \ref{tab:Sample}). The high-ionisation [NeV] and [NeVI] lines also predominantly trace the disk kinematics, with a kinematic axis consistent with [NeII], unlike in Seyfert galaxies \citep[e.g.][]{HermosaMunoz2024, Zhang2024b, Veenema2026, Donnan2026}, where there is a clear separation of outflow, traced by [NeVI], and disk, traced by [NeII]. This is particularly clear for J1509. J1100 is an exception and does show the typical behaviour, where [NeV] and [NeVI] trace a different velocity structure than [NeII]. 

The velocity maps of [NeV] and [NeVI] show disk kinematics in the QSO2s. This suggests that the AGN, of higher luminosity than the Seyfert galaxies, has ionised all the surrounding gas \citep{Bianchin2026}. From spatially resolved Baldwin, Phillips, and Terlevich (BPT) diagrams involving optical emission lines detected with VLT/MUSE, \citet{Venturi2023} and \citet{Ulivi2024} showed that the bulk of the line-emitting gas in QSO2s, including J1010, J1100, and J1430, is photoionised by the AGN. In contrast, the disks of Seyfert galaxies include a significant fraction of the gas, which is sometimes dominant and is ionised by star formation \citep{Mingozzi2019}. 

As mentioned in Sect. \ref{sec:Data}, the five QSO2s we studied have ionised and molecular outflows that have been characterised using different methods \citep{Ramos2017, RamosAlmeida2019, RamosAlmeida2022, Speranza24, Audibert25}. We show the velocity dispersion for [NeV] in J1509 and J1430 in Fig. \ref{fig:J1509Sigma}. We created this map by fitting a Gaussian line profile to each spaxel and fitting for the width of the Gaussian as parametrised by the velocity dispersion, $\sigma$, and the line spread function of MIRI MRS at the observed wavelength. The [NeV] emission shows enhanced velocity dispersion along the outflow for J1509 and J1430, while the dispersion is low across the disk. Therefore, the outflow is indeed present in the highly ionised gas, but only apparent in the velocity dispersion for J1509 and J1430 \footnote{We note that the ionised outflow in these QSO2s is detected in velocity and velocity dispersion using parametric and non-parametric methods to fit the emission lines \citep{RamosAlmeida2019, Speranza24}.}. This suggests that the ionised gas, unlike the molecular and dust phases, is highly turbulent and is thus disturbed by the outflow, as previously shown from a non-parametric analysis of the optical [OIII] emission of J1509 \citep{Bessiere24,Speranza24}. For J1100, the outflow can be seen in the velocity map, where the position angle of the kinematics is different for high-ionisation lines compared to the low-ionisation lines (such as [NeII]).

\subsubsection{Molecular gas}
The velocity maps of the molecular gas as traced by the rotational H$_2$ S(5), S(3), and S(1) lines broadly trace rotation with a position angle consistent with the galaxy disk. From these maps alone, any outflow contribution is not immediately clear. However, we investigate the presence of the warm molecular outflows further in Sect. \ref{sec:DiskSub}. J1100 is an exception, where the higher J transitions such as S(3) are more similar to the high-ionisation lines and the 11.3 $\mu$m PAH.

\subsubsection{PAHs}
\begin{figure}        
    \centering
        \includegraphics[width=\columnwidth]{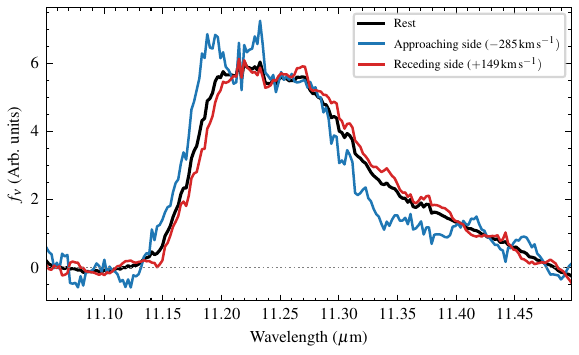}
    \caption{Profiles of the 11.3 $\mu$m PAH feature of the rest-frame emission (eigenspectrum of the first component), the blue- and redshifted sides of the outflow structure with velocities displayed in the legend. These are the average profiles within 0.4'' radius apertures.}
    \label{fig:Profile} 
\end{figure}

Out of the five objects, J1509, J1430, and J1100 show coherent kinematic structures in the 11.3 $\mu$m PAH feature. In all three cases, there are clear redshifted and blueshifted structures with different orientation that the disk, which might suggest the presence of the outflow in the PAH velocity maps. We show the profile of the 11.3 $\mu$m PAH feature for J1509 in Fig. \ref{fig:Profile}, where we show the rest-frame profile and the profiles of the redshifted and blueshifted lobes.

Due to the location of the 11.3 $\mu$m PAH feature within the 9.7 $\mu$m silicate feature, we considered whether extinction is an issue that might cause erroneous velocity measurements. If there is a strong change in extinction spatially, in regions of higher extinction the short-wavelength side of the feature is suppressed relative to the longer wavelength side, mimicking a redshift. This can result in redder velocities when the extinction is high $\Delta\tau_{9.7}\gtrsim1$. While the nuclear spectra of our sample can show deep silicate features \citep{RamosAlmeida2025}, the circumnuclear emission in our PSF-subtracted cubes does not show high levels of extinction (see Fig. \ref{fig:Spec}). Moreover, fitting the circumnuclear spectra with SPIRIT \citep{Donnan2024} \footnote{\url{https://github.com/FergusDonnan/SPIRIT}} yielded low extinction measurements, with an H$_2$ extinction of $\tau_{9.7}\sim0.4$ for J1509. Therefore, we conclude that extinction does not affect our measured velocities in any significant way.

As mentioned above, only J1100 shows [NeV] and [NeVI] velocity maps that are clearly distinct from the disk rotation. The 11.3 $\mu$m PAH velocity map indeed resembles the [NeV] much more than [NeII] in this target, suggesting that the 11.3 $\mu$m PAH is dominated by the outflow in this source. We explore the PAH velocity maps further, after subtracting the disk kinematics, in Sect. \ref{sec:DiskSub}.

We find it challenging to construct velocity maps for the 6.2 $\mu$m PAH because the signal-to-noise ratio of this feature is low. The only target with a sufficient signal-to-noise ratio to detect a coherent signal in the second principal component was J1509. In this target, however, the PCA detected a different intrinsic profile rather than a Doppler shift. This was also found in \citet{Donnan2026}, and so the PCA detects differences in the intrinsic profile of the 6.2 $\mu$m PAH feature rather than velocity shifts. As the 6.2 $\mu$m PAH traces ionised PAHs while the 11.3 $\mu$m feature traces predominantly neutral PAHs, the issues with the 6.2 $\mu$m reflect what has been found in previous studies, where AGN outflows have PAHs that are observed to be be more neutral than those in star-forming regions \citep[][]{ODowd2009, Diamond-Stanic2010, Garcia-Bernete2022b, Garcia-Bernete2022c, Garcia-Bernete2024b, Rigopoulou2024, Zhang2024, Donnelly2024, RamosAlmeida2025}. Studies of the profile of the 6.2 $\mu$m PAH feature suggested that the differences in the intrinsic profile arise from the prominence of a sub-feature at $\sim6.35\mu$m that might be due to a different molecular species \citep{Peeters2002, Canelo2026}. As we once again struggled to extract any kinematics of the 6.2 $\mu$m PAH due to differences in the intrinsic profile and its relatively low brightness compared to the 11.3 $\mu$m feature (\textcolor{blue}{Ramos Almeida in prep.}), this re-enforces the idea that the ionised PAHs are strongly affected in AGN outflows.

\begin{figure}        
    \centering
        \includegraphics[width=\columnwidth]{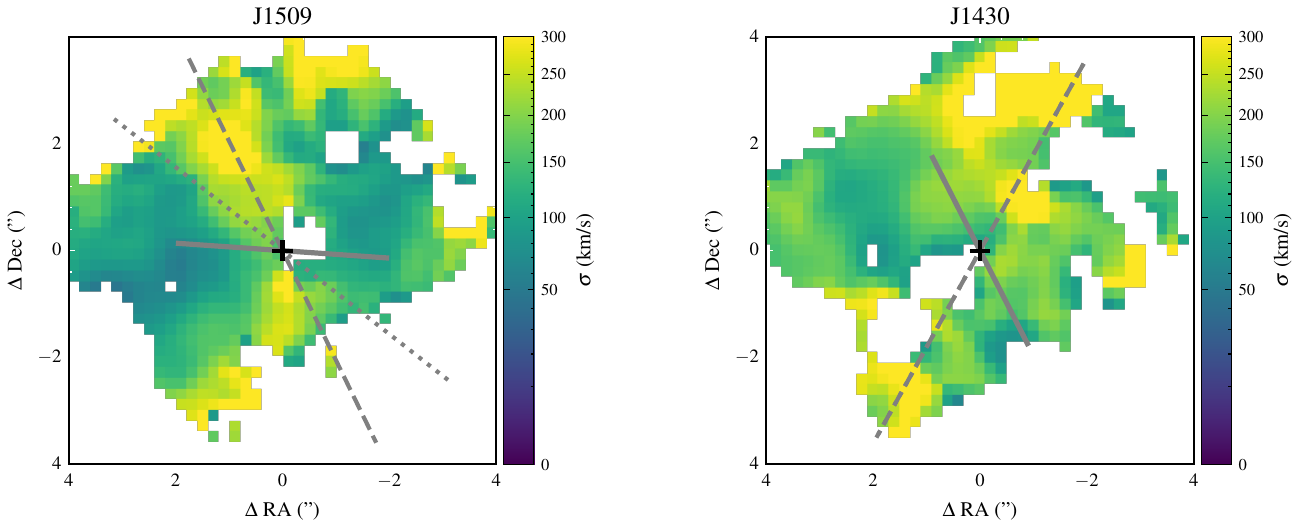}
    \caption{Velocity dispersion, $\sigma$, of [NeV] for J1509 and J1430, as measured by fitting a Gaussian line profiles to each spaxel. The AGN position is shown with the black cross, and the position angle of the galaxy disk is shown with the solid grey line. The dotted and dashed lines show the position angles of the PAH outflow and H$_2$ outflow for 
    J1509 (see Fig. \ref{fig:J1509} for reference). For J1430, the PAH and H$_2$ outflow are shown by the dashed line. The solid grey line shows the position angle of the disk.}
    \label{fig:J1509Sigma} 
\end{figure}

\subsection{Disk subtraction}
\label{sec:DiskSub}
As described in Sect. \ref{sec:DiskModelling}, we modelled the galaxy disk velocities, fixing the orientation based on CO constraints \citep[][]{RamosAlmeida2022, Audibert25}. By subtracting the disk velocities from the observed 2D velocity maps presented in Fig. \ref{fig:VelMaps}, we revealed any residual non-circular velocities that might stem from the AGN-driven outflows.

\begin{figure}
        \includegraphics[width=\columnwidth]{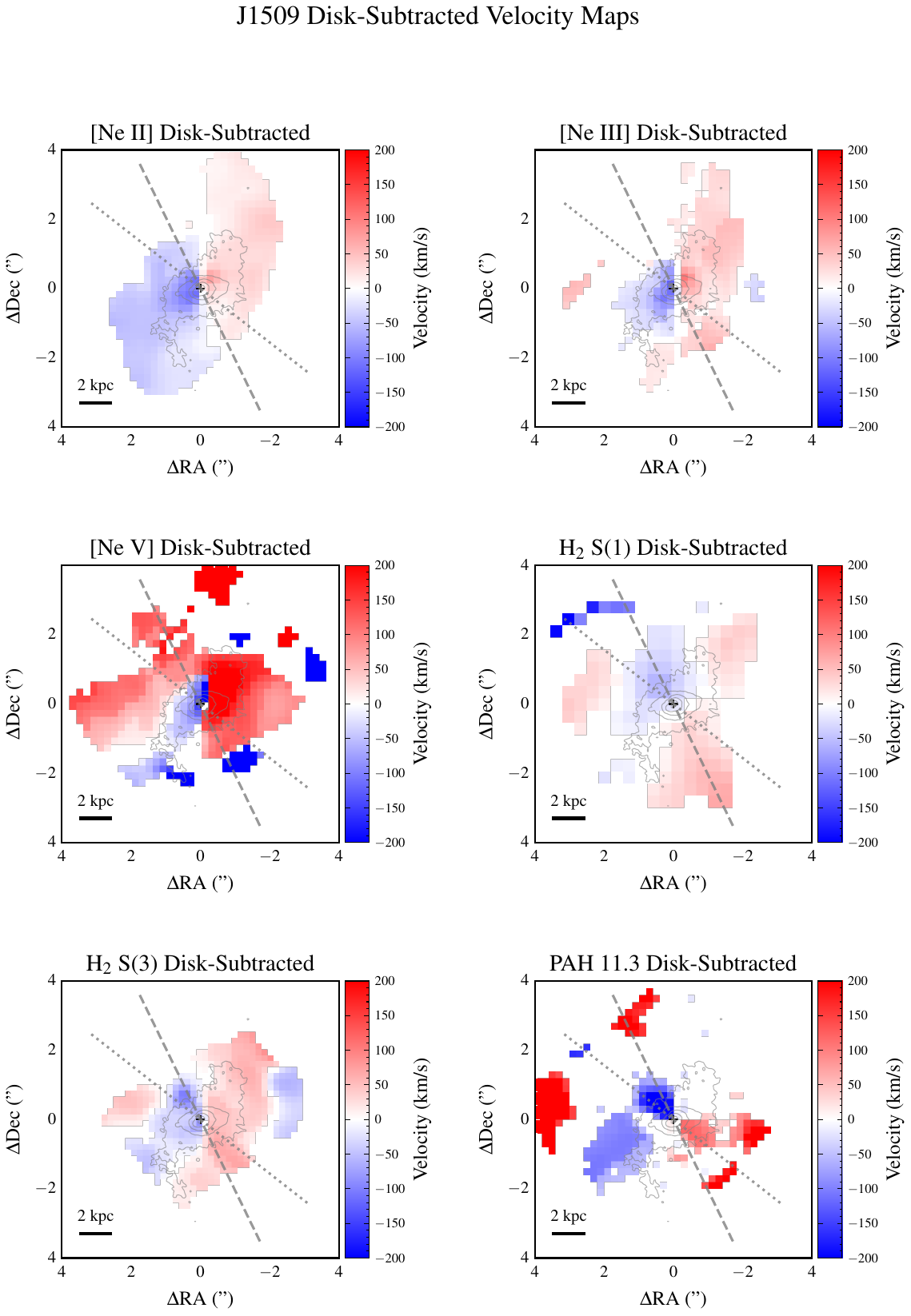}
    \caption{Velocity maps for J1509 after subtracting a disk model. The original velocity maps are shown in Fig. \ref{fig:VelMaps}. The grey contours show the CO(2-1) moment 0 map from \citet{RamosAlmeida2022}. We show the measured position angle of the H$_2$ outflow with the dashed grey line, and the position angle of the PAH outflow is shown with the dotted line. We only show spaxels where the velocity is $> 1\sigma$, where $\sigma$ is the error in the velocity.}
    \label{fig:J1509} 
\end{figure}

\begin{figure}
        \includegraphics[width=\columnwidth]{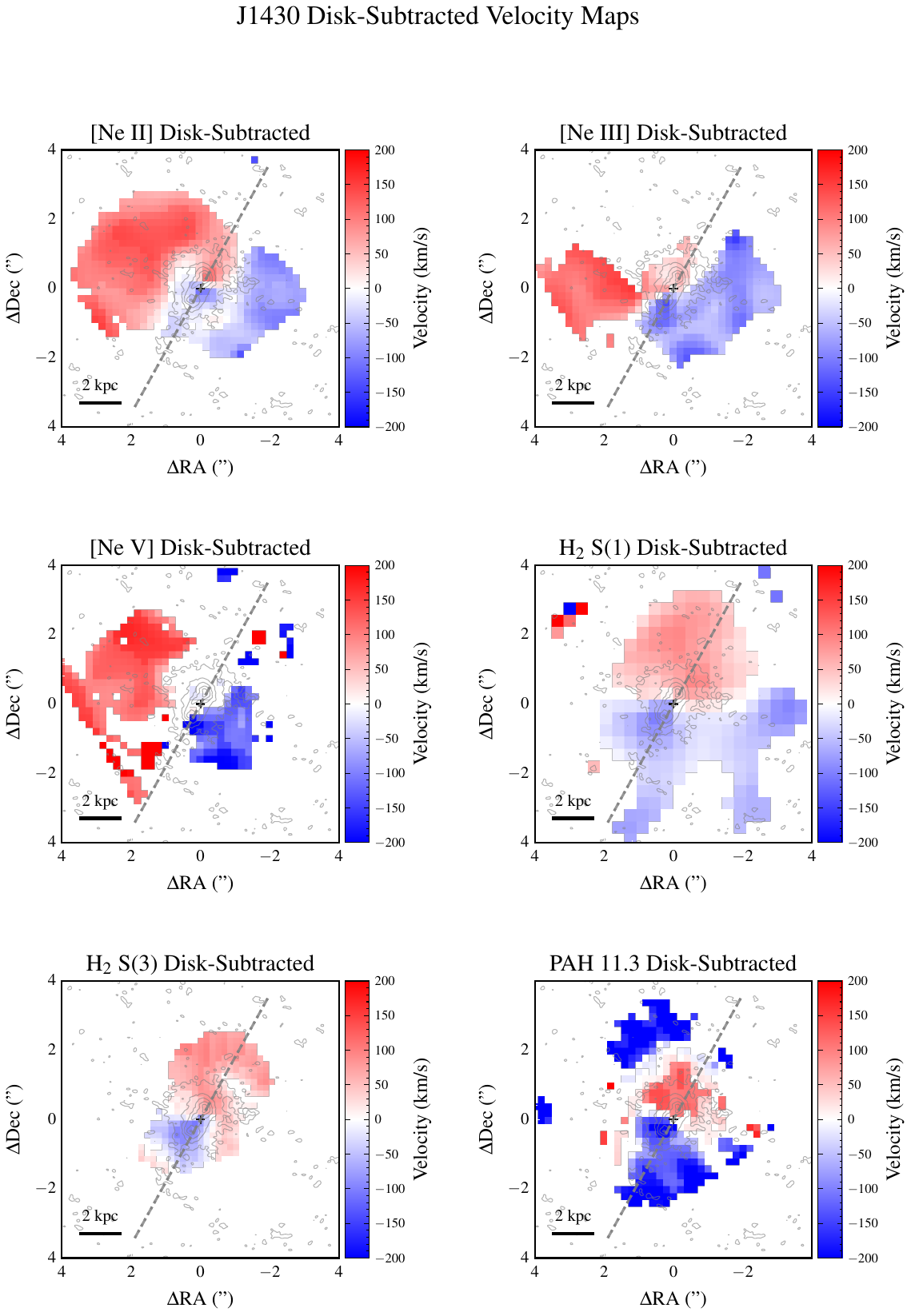}
    \caption{Same as Fig. \ref{fig:J1509} but for J1430. Here, the position angle of the PAH and H$_2$ outflows match, and a single dashed line is therefore plotted to show this direction.}
    \label{fig:J1430} 
\end{figure}

\begin{figure}
        \includegraphics[width=\columnwidth]{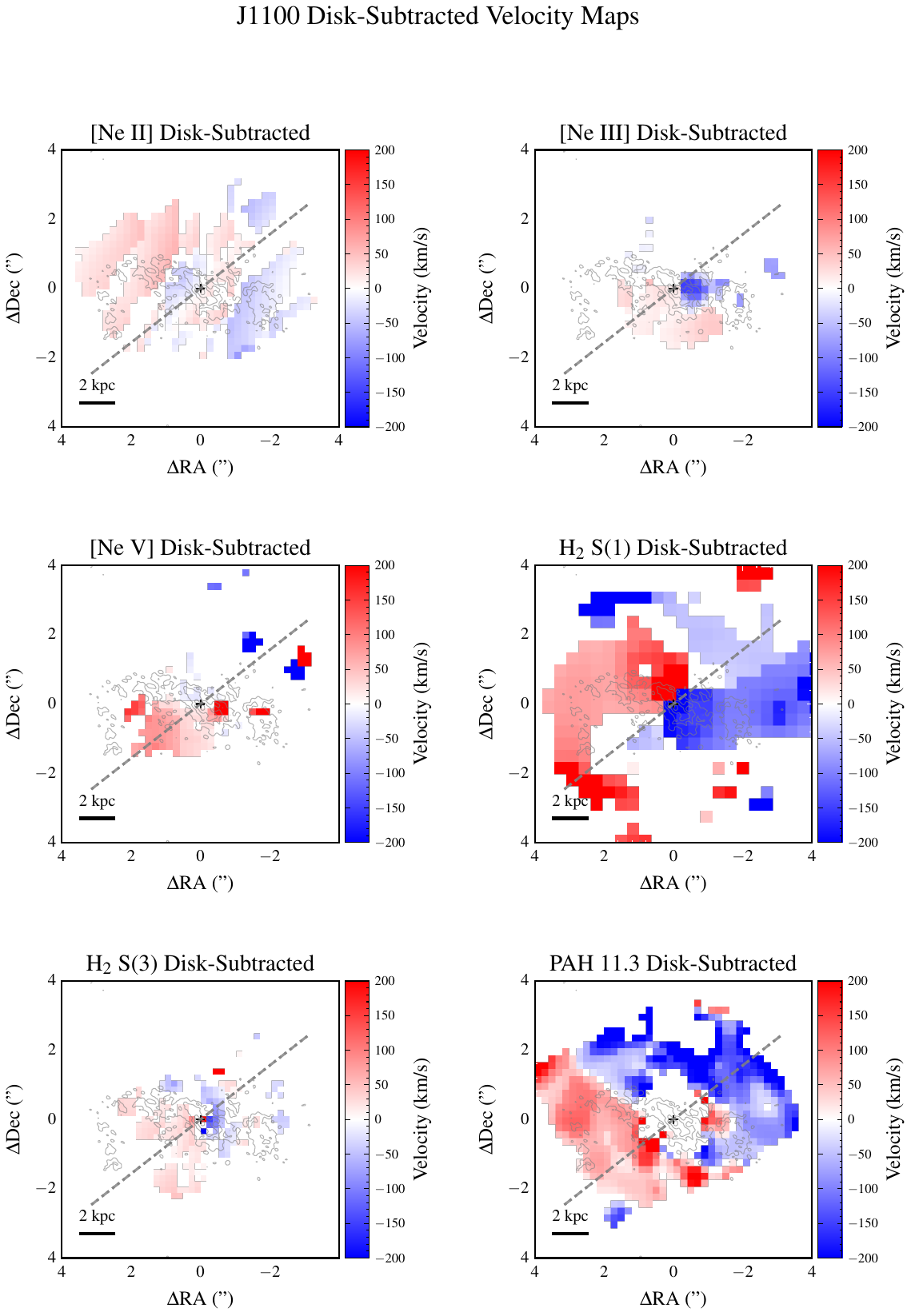}
    \caption{Same as Fig. \ref{fig:J1509} but for J1100. Here, the position angle of the  PAH and H$_2$ outflows match, and a single dashed line is therefore plotted to show this direction.}
    \label{fig:J1100} 
\end{figure}

We show the disk-subtracted velocity maps for the PAH features in Figs. \ref{fig:J1509}, \ref{fig:J1430}, and \ref{fig:J1100} for J1509, J1430, and J1100, respectively. We focus on these three targets as they show a clear detection of kinematics in the 11.3 $\mu$m PAH feature. We measured the position angles of the outflow structures by fitting a straight line to the peak of the velocity structures, as described in the following sections. These angles are shown in Table \ref{tab:PAs_checked}.

\subsubsection{J1509}
J1509 is a barred spiral galaxy, and the bar and spiral arms are both detected in CO(2-1) \citep{RamosAlmeida2022, Audibert25}. Gas outflows have been reported in low-ionisation atomic gas \citep{Bessiere24,Speranza24}, high-ionisation atomic and warm H$_2$ gas \citep{RamosAlmeida2019}, and in CO \citep{RamosAlmeida2022,Audibert25}. The ionised outflow, traced by [OIII], [SiVI], and Pa$\alpha$, only shows the blueshifted component ($\rm PA\sim -40^{\circ}=320^{\circ}$; \citealt{Speranza24}), but the CO outflow shows blue- and redshifted emission to the north and south, respectively \citep{RamosAlmeida2022}. We summarise the orientations of the PAHs and H$_2$ in comparison with the literature in Table \ref{tab:PAs_checked}.
The latter authors reported that because of the bar, the CO kinematics suggest a competition between inflowing and outflowing gas.

The residual velocity map of the 11.3 $\mu$m PAH for J1509 (see Fig. \ref{fig:J1509}) shows clear blue- and redshifted cone structures to the north-east and south-west, respectively, and is perpendicular to the bar (as traced by the CO emission) in the inner regions of the galaxy. We measure the position angle of this structure to be 232$^{\circ}$ towards the redshifted side, where 0$^{\circ}$ is north, and the position angle increases to the east (left). A similar structure can also be seen in the H$_2$ S(1) line, also perpendicular to the bar, as shown in Fig. \ref{fig:J1509}. This appears at a slightly different position angle of 206$^{\circ}$, however. We show these angles in Fig. \ref{fig:J1509} and Table \ref{tab:PAs_checked}. The H$_2$ S(3) line also shows this structure, with a position angle closer to the PAHs than the S(1) line. The bar can also be seen in the disk-subtracted velocity of the S(3) line. 

We thus detect non-circular motions related to an outflow that is present in the dust and molecular gas phase, but it is not seen in either of the velocity maps of the ionised gas from [NeII] and [NeV]. Nevertheless, as mentioned in Sect. \ref{sec:velocity_maps}, we indicate the outflow in the velocity dispersion of [NeV], which we show in Fig. \ref{fig:J1509Sigma}. We find that the outflow regions to the north and south show a significantly higher velocity dispersion than the galaxy disk. This suggests that the ionised gas is heavily disrupted by the outflow, making its signal in the velocity maps derived from PCA decomposition weak but apparent in the velocity dispersion map.

In addition to the potential outflow seen in the 11.3 $\mu$m PAH, H$_2$ S(1), and H$_2$ S(3) lines, Fig. \ref{fig:J1509} shows red- and blueshifted structure to the north-west and south-east, respectively, in the H$_2$ residual velocity map and tentatively in the PAH residual velocity map. These residuals are consistent with the ionised gas kinematics as traced by [NeII] and [NeV] and also with the spiral arms detected in CO (see Fig. 13 in \citealt{RamosAlmeida2022}). Because these velocity structures are consistent with the spiral arms, we suggest that they trace an inflow to the nucleus. Moreover, the spiral arm regions show the lowest velocity dispersion in the ionised gas, as shown in Fig. \ref{fig:J1509Sigma}.

\begin{table}[h]
  \caption{Position angles of the PAH and H$_2$ outflows detected from JWST compared to the literature.}
\centering
\renewcommand{\arraystretch}{1.5}
\setlength{\tabcolsep}{3pt}
\label{tab:PAs_checked}
\begin{tabular}{l|cc|cc|cc|cc}
\hline
\multirow{2}{*}{Name} & \multicolumn{2}{c|}{PAH} & \multicolumn{2}{c|}{H$_2$} & \multicolumn{2}{c|}{[OIII]} & \multicolumn{2}{c}{CO} \\
 & Red & Blue & Red & Blue & Red & Blue & Red & Blue \\
\hline
J1509 & 232$^{\circ}$ & 52$^{\circ}$ & 206$^{\circ}$ & 26$^{\circ}$& \dots & -40$^{\circ}$ & 132$^{\circ}$ & -8$^{\circ}$ \\
J1430 & -29$^{\circ}$ & 151$^{\circ}$ & -29$^{\circ}$ & 151$^{\circ}$ & 198$^{\circ}$ & 65$^{\circ}$ & -18$^{\circ}$ & 188$^{\circ}$ \\
J1100 & 128$^{\circ}$ & -52$^{\circ}$ & 128$^{\circ}$ & -52$^{\circ}$ & \dots  & 63$^{\circ}$ &55$^{\circ}$ &200$^{\circ}$ \\
\hline
\end{tabular}
\tablefoot{Position angles are measured from north to east. [OIII] values are from \citet{Speranza24} and CO values from \citet{RamosAlmeida2022} in the case of J1509 and from \citet{Audibert25} for J1430 and J1100.}
\end{table}

\subsubsection{J1430}
J1430 is a post-merger galaxy with a $\sim$55\% contribution of disk rotation in CO(2-1), and the rest is dispersion dominated \citep{Audibert23}. For this reason, we inferred the position angle of the disk from the [NeII] velocity map, as shown in Fig. \ref{fig:VelMaps}, of $\sim 27^{\circ}$ rather than $\sim 3^{\circ}$ from \citet{Audibert25}, which in the other four galaxies traces the disk very well. There are still some residuals in the disk-subtracted [NeII] velocity map, which might suggest that there is a warp in the disk where the position angle changes with radius. The disk-subtracted velocity maps are shown in Fig. \ref{fig:J1430}. J1430 also has a multi-phase outflow detected in low- and high-ionisation atomic gas \citep{Harrison2014,Harrison2015,Ramos2017,Venturi2023,Bessiere24,Speranza24}, warm molecular gas \citep{Zanchettin2025}, and CO \citep{RamosAlmeida2022,Audibert23,Audibert25}. We show the position angles of these outflows in Table \ref{tab:PAs_checked}.

The residual H$_2$ S(1) and S(3) velocity maps in Fig. \ref{fig:J1430} shows the outflow, matching the position angle of the CO(2-1) emission (the disk PA and the outflow). We measured this position angle to be -29$^{\circ}$ to the redshifted side measured from the north.

The 11.3 $\mu$m PAH velocity map receives little contribution from the disk, with the red- and blueshifted sides of the velocity field matching that of the molecular outflow and the position angle CO(2-1) contours even without disk subtraction (see Fig. \ref{fig:VelMaps}). 

Similar to J1509, we did not detect the outflow in ionised gas using the velocity maps generated from PCA decomposition. The velocity dispersion of the ionised gas along the outflow is high, however, as shown in Fig. \ref{fig:J1509Sigma}. Previous studies \citep[][]{Ramos2017, Speranza24} reported the detection of an ionised outflow at a position angle $\sim65^{\circ}-70^{\circ}$, which matches the direction of a jet detected in the radio \citep{Harrison2015}. This places the ionised outflow and the jet perpendicular to the outflow seen in the molecular gas and the PAHs, where \citet{Audibert23} suggested that the jet inflates bubbles of molecular gas, driving a lateral molecular outflow.

\subsubsection{J1100}

J1100 is a barred spiral galaxy with an apparently undisturbed morphology \citep{Fischer2018}. The disk-subtracted velocity maps are shown in Fig. \ref{fig:J1100}. As in the case of J1509, the bar and the spiral arms are detected in CO \citep{RamosAlmeida2022,Audibert25}. The bar has a radius of $\sim$2.5''~and PA$\sim$85$^{\circ}$, and it is connected with the inner part of the spiral arms (see Fig. 7 in \citealt{RamosAlmeida2022}). The H$_2$ S(1) velocity map shown in Fig. \ref{fig:VelMaps} resembles the morphology and kinematics of the CO(2-1). Gas outflows have been reported in ionised gas \citep{Harrison2014,Fischer2018,Bessiere24,Speranza24,Ulivi2024} and in CO \citep{RamosAlmeida2022,Audibert25}, with the latter having position angles of 200$^{\circ}$ and 55$^{\circ}$ for its blue- and redshifted sides (see Table \ref{tab:PAs_checked}).

The disk kinematics contribute little to the velocity map of the 11.3 $\mu$m PAH, where the position angle of the kinematic axis matches that of the outflow as seen by [NeV], and therefore, the presence of the dusty outflow is clear in J1100. Moreover, the kinematics of the disk-subtracted [NeIII] line matches that of [NeV]. We measured a position angle of 128$^{\circ}$ for the outflow in J1100. We note that the ionised outflow orientation measured here from the neon maps does not match that of the [OIII] outflow reported by \citet{Speranza24}, as shown in Table \ref{tab:PAs_checked}. This might be due to the limited angular resolution of the GTC/MEGARA observations of this QSO2 (full width at half maximum of $\sim$1.2 arcsec).  VLT/MUSE observations of this target indeed show an enhancement in the velocity dispersion to the north-west (\citealt{Ulivi2024} and Bianchin et al. in prep.).

The [NeII] velocity map matches that of the disk very well (as seen in Fig. \ref{fig:VelMaps}), where the disk-subtracted map shows very little residuals with the mean residuals $\lesssim10$ km s$^{-1}$.

The H$_2$ S(1) map is more complex, with the disk-subtracted velocity map showing multiple components. In the inner regions lies %and inner disk 
the bar, extending from the north-east to the south-west, with a position angle consistent with the CO(2-1) emission \citep{RamosAlmeida2022, Audibert25}. 
The outer regions resemble the spiral arms seen in CO. 
There might also be an outflow contribution in the molecular gas, where the S(5) and S(3) line, as shown in Fig. \ref{fig:VelMaps}, shows a velocity map more consistent with [NeV]. The increase in the contribution of the outflow in the higher J H$_2$ lines suggests that the molecular gas is hot in the outflow \citep[e.g.][]{Davies2024}, appearing in the higher J rotational transitions of H$_2$.

\section{Discussion}
\label{sec:Discussion}

\subsection{How common are dusty outflows}
We found evidence of PAHs in the outflow of three out of the five QSO2s in our sample. %however 
Moreover, out of those that actually show PAH kinematics, we detected kinematics consistent with an outflow in all three. The signal-to-noise ratio of the remaining two objects, J1356 and J1010, is insufficient to produce PAH velocity maps (see \ref{fig:VelMaps}). For the case of J1356, this may be because this system is a major merger, and thus, the kinematics are simply too complex to show a coherent velocity map, while J1010 shows very little PAH emission, however.

\citet{Donnan2026} found two objects with PAH kinematic detections of the outflow in a sample of Seyfert galaxies. The full sample included ten objects, but no PAHs were determined in the outflows for five objects for a multitude of reasons. Some the galaxies were edge-on, and thus, the projected outflow velocities were extremely low simply due to orientation, and therefore could not be detected. The signal-to-noise ratio of other objects was insufficient to produce PAH velocity maps, similar to J1356 and J1010. Therefore, two out of five Seyfert galaxies showed PAH outflow kinematics. Moreover, the three objects for which PAH outflow kinematics were ruled out had lower Eddington ratios than the two that showed PAH outflow kinematics. 

\begin{figure}
        \includegraphics[width=\columnwidth]{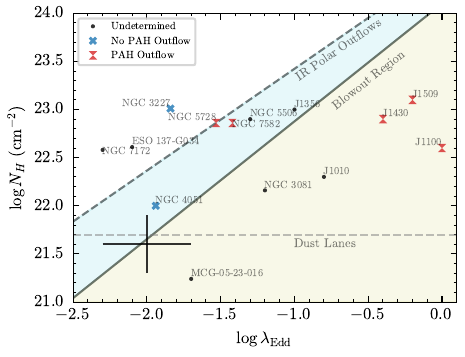}
    \caption{Diagram of the column density, $N_H$, vs. the Eddington ratio, $\lambda_{\textrm{Edd}}$ of the five QSO2s in this work and the Seyfert galaxies in \citet{Donnan2026}. The sources for which we detect outflowing PAHs are shown with the red hourglass marker, and sources for which we find no dusty outflows via PAH kinematics are shown with the blue cross. The sources for which this was neither confirmed nor ruled out are shown with the black dots. We show the IR polar outflow region from \citet{Venanzi2020} as well as the blowout region from \citet{Fabian2008}, where we theoretically expect dusty polar outflows. At column densities $<5\times10^{21}$ cm$^{-1}$, dust lanes can cause the obscuration \citep{Fabian2006, Ricci2017b}, which we show with the horizontal dashed grey line.}
    \label{fig:EddPlot} 
\end{figure}

The potential dependence of dusty outflows on the Eddington ratio was predicted by \citet[][]{Fabian2008}, where the position of the object in a plot of the column density, $N_{\rm H}$, versus Eddington ratio, $\lambda_{\rm Edd}$, indicates when a dusty outflow is expected. We show this in Fig. \ref{fig:EddPlot}.

The column densities were calculated from CO observations \citep{Garcia-Burillo2024} assuming Milky Way values of $\alpha_{\rm CO}$ and CO ratio scaling factors derived from high-resolution data of NGC 1068 \citep{Garcia-Burillo2014, Viti2014}. Due to the assumptions for this calculation and the determination of the Eddington ratio, we show a large error range displayed in Fig. \ref{fig:EddPlot}. As NGC\,4051 lacks CO data, we instead used the X-ray measured column density from \citet{Pounds2004}. The final source from \citet{Donnan2026} that is missing from this plot is ESO 420-G013, which lacks any reliable column density or Eddington ratios measured in the literature. However, this is one of the sources for which a dusty outflow was ruled out after no evidence of the outflow in the 11.3 $\mu$m PAH velocity map was detected.

While the sample size is small, Fig. \ref{fig:EddPlot} shows that all the sources with dusty outflows (red hourglass marker) tend to have higher Eddington ratios than the two sources for which dusty outflows were ruled out,  namely NGC\,4051 and NGC\,3227 (blue cross marker). The remaining sources we designate as unknown, either due to a lack of signal-to-noise ratio or projection effects, where the outflow was edge-on, and thus, the velocity component could not be measured. 

All the sources for which dusty outflows were observed appear in either the IR polar outflow region from \citet{Venanzi2020} or in the blowout region from \citet{Fabian2008} in the parameter space, where dusty outflows are theoretically expected. As previously reported by \citet{Ramos2026}, the five QSO2s are in either the IR polar outflow or blowout region in Fig. \ref{fig:EddPlot}, and the nuclear infrared spectrum of J1010 was successfully reproduced with the CAT3D-wind torus model of \citet{Honig2017}. It is therefore very likely that the five QSO2s have dusty outflows, but the current JWST observations do not allow us to detect PAH kinematics in J1010 and J1356, as previously mentioned. We also note that one of the sources for which dusty outflows were ruled out, NGC 4051, appears in the IR polar outflow region, but its column density is more uncertain due to the reasoning above.

\subsection{Acceleration of dust grains}
\begin{figure}
        \includegraphics[width=\columnwidth]{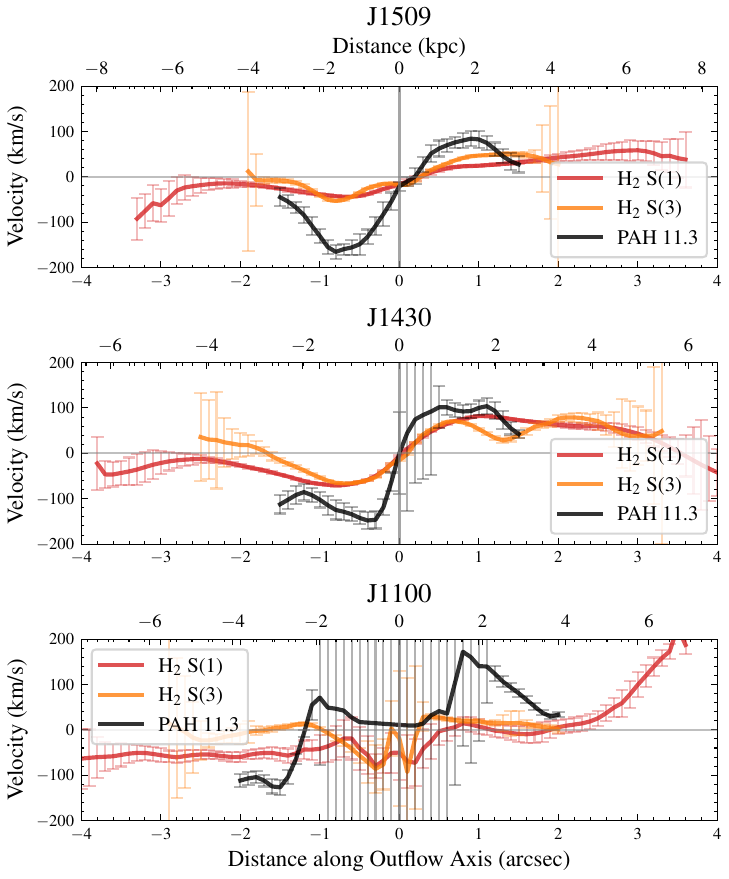}
    \caption{Velocity profiles of the 11.3 $\mu$m PAH, H$_2$ S(3), and H$_2$ S(1) line along the outflow axis for J1509, J1430, and J1100. The profiles were extracted from the disk-subtracted maps. The top axes show the distance in kpc.}
    \label{fig:VelProf} 
\end{figure}

To investigate the mechanisms behind the different phases of the outflow, we extracted the velocity profile along the outflow axis for the 11.3 $\mu$m PAH, the H$_2$ S(3), and H$_2$ S(1) features from the disk subtracted velocity maps, through an aperture of 1 arcsecond in increments of 0.1 arcsecond. This is shown in Fig. \ref{fig:VelProf}.

The 11.3 $\mu$m PAH shows a higher acceleration and then higher deceleration than the H$_2$ S(1) and S(3) lines. Moreover, the PAHs reach a higher peak velocity before decelerating. This matches what was observed in Seyfert galaxies in \citet{Donnan2026}, although it is worth noting that the spatial scales are greater here. The peak velocity is reached at $\sim1-2$ kpc, while the outflows detected in \citet{Donnan2026} are on sub-kiloparsec scales. This likely reflects the higher luminosity of the targets in this work, which are known to drive larger outflows than lower-luminosity AGN \citep[e.g.][]{Kang2018, Kakkad2022, Kim2023, Harrison2024}.

One interpretation of the higher acceleration is that PAH grains are easier to accelerate due to their larger absorption cross-section relative to mass compared to H$_2$, which  leads to a stronger radiative acceleration for the PAHs \citep[e.g.][]{Tielens2010}. Similarly, PAH molecules might also decelerate faster when the outflow sweeps up material in the circumnuclear environment because PAHs have a large geometric cross-section per mass. This scenario implies a decoupling of the PAHs from the molecular gas that is difficult to explain considering that the timescale on which the PAHs equilibrate with the gas \citep[e.g.][]{Li2001} is much lower than the outflow timescale of $\sim10^6$ yr. Perhaps the molecular gas coupled to the PAHs simply does not emit.

Shocks can also play a role in these outflows as they are thought to destroy PAHs when they reach $\gtrsim100$ km s$^{-1}$, and so only slowly moving PAHs can survive \citep[e.g.][]{Micelotta2010}. This might explain why we observe no PAHs exceeding $\gtrsim100$ km s$^{-1}$, although a more thorough analysis is needed to determine the presence of shocks.

This result suggests that the dusty outflow phase of AGN outflows might be short lived, where either the grains decelerate or are destroyed before the molecular or ionised gas. If they do indeed decelerate, they may fall back towards the AGN. This scenario has been postulated by \citet{Wada2012}, who proposed a fountain model in which dust grains that fail to be ejected fall back towards the AGN. The authors further suggested that the dusty torus itself is the product of these failed winds. We cannot conclusively confirm this here, but the fountain-model scenario is consistent with our observations.

\section{Conclusions}
We have applied PCA tomography to measure the kinematics of PAHs in five QSO2s. Three out of the five objects show a clear detection of kinematics in the 11.3 $\mu$m PAH feature, namely J1509, J1430 and J1100. Our main findings are listed below.
\begin{itemize}

    \item In the QSO2s, high-ionisation lines of [NeV] and [NeVI] trace gas rotation in all cases except J1100, unlike in Seyfert galaxies, where they trace outflowing gas. This is likely due to the higher AGN luminosities of the quasars, which photo-ionise the bulk of the gas in the galaxy disks.

    \item We observed two structures in the 11.3 $\mu$m PAH velocity map for J1509, particularly after subtracting the disk kinematics. An inflow structure is coincident with spiral arms, and this was also detected in molecular and ionised gas via the H$_2$ S(1) and [NeII] lines, respectively. Clear outflow kinematics are also clear in the 11.3 $\mu$m PAH, which is also present in the kinematics of the H$_2$ S(1) line and a region of high velocity dispersion in highly ionised gas in [NeV].

    \item In J1430, the kinematics of the 11.3 $\mu$m PAH are dominated by the outflow, which matches that of the H$_2$ after subtracting the disk kinematics. The outflow is consistent with the CO(2-1) emission on J1430.

    \item In J1100, we also see kinematic evidence of the outflow in the 11.3 $\mu$m PAH, consistent with [NeV] and the higher J H$_2$ S(5) and S(3) lines. The H$_2$ S(1) line is consistent with CO(2-1) and is dominated by the bar and innermost part of the spiral arms. 

    \item We detected kinematic evidence of PAHs in AGN outflows in all QSO2s for which velocity maps could be constructed. This suggests that dusty outflows are more common in quasars than in Seyfert galaxies, which occupy a parameter space with higher luminosities and Eddington ratios and similar column densities than Seyfert galaxies. 

    \item As with the Seyfert galaxies, the orientation of the PAH outflows matches that of the molecular gas, although the velocities are different. 
    
    \item The acceleration of the PAHs in the outflow is higher than the molecular gas outflows, consistent with what was seen in the Seyfert galaxies. This might reflect the higher ratio of the surface area to mass  of PAHs and/or dust grains compared to H$_2$, making them easier to radiatively accelerate by UV photons.
    
\end{itemize}
This work, combined with the lower luminosity sample presented in \citet{Donnan2026}, shows that dusty outflows may be common in AGN with high Eddington ratios. Therefore, dusty outflows are a viable mechanism for AGN to clear their dust, potentially moving from dust-obscured phases such as hot dust-obscured galaxies to low-dust AGN such as little red dots in the high-redshift Universe.

\begin{acknowledgements}
We thank the referee for their feedback and review of the paper.
FRD and KS acknowledges funding support from grant JWST-GO-05279.002. 
CRA, AA, and MB thank the Agencia Estatal de Investigación of the Ministerio de Ciencia, Innovación y Universidades (MCIU/AEI) under the grant “Tracking active galactic nuclei feedback from parsec to kiloparsec scales”, with reference PID2022-141105NB-I00 and the European Regional Development Fund (ERDF). O..G.-M. acknowledges the support received by the UNAM DGAPA-PAPIIT project IN109123, SEHCITI Cien- cia de Frontera project CF-2023-G100, and UNAM DGAPA sabbatical grants. MPS acknowledges support from grants RYC2021-033094-I, CNS2023-145506, and PID2023-146667NB-I00 funded by MCIN/AEI/10.13039/501100011033 and the European Union NextGenerationEU/PRTR. IGB is supported by the Programa Atracci\'on de Talento Investigador ``C\'esar Nombela'' via grant 2023-T1/TEC-29030 funded by the Community of Madrid, and acknowledges support from the research project PID2024-159902NA-I00 funded by the Spanish Ministry of Science and Innovation/State Agency of Research (MCIN/AEI/10.13039/501100011033) and FSE+. MB acknowledges support from the Juan de La Cierva scholarship with reference JDC2023-052684-I, funded by MICIU/AEI/10.13039/501100011033 and FSE+. 
\end{acknowledgements}

\bibliography{References}{}

@ARTICLE{Wada2012,
       author = {{Wada}, Keiichi},
        title = "{Radiation-driven Fountain and Origin of Torus around Active Galactic Nuclei}",
      journal = {\apj},
         year = 2012,
        month = oct,
       volume = {758},
       number = {1},
          eid = {66},
        pages = {66},
          doi = {10.1088/0004-637X/758/1/66},
archivePrefix = {arXiv},
       eprint = {1208.5272},
 primaryClass = {astro-ph.GA},
       adsurl = {https://ui.adsabs.harvard.edu/abs/2012ApJ...758...66W}
}

@ARTICLE{Phan2019,
       author = {{Phan}, Du and {Pradhan}, Neeraj and {Jankowiak}, Martin},
        title = "{Composable Effects for Flexible and Accelerated Probabilistic Programming in NumPyro}",
      journal = {arXiv e-prints},
         year = 2019,
        month = dec,
          eid = {arXiv:1912.11554},
        pages = {arXiv:1912.11554},
archivePrefix = {arXiv},
       eprint = {1912.11554},
 primaryClass = {stat.ML},
       adsurl = {https://ui.adsabs.harvard.edu/abs/2019arXiv191211554P}
}

@ARTICLE{Garcia-Burillo2014,
       author = {{Garc{\'\i}a-Burillo}, S. and {Combes}, F. and {Usero}, A. and {Aalto}, S. and {Krips}, M. and {Viti}, S. and {Alonso-Herrero}, A. and {Hunt}, L.~K. and {Schinnerer}, E. and {Baker}, A.~J. and {Boone}, F. and {Casasola}, V. and {Colina}, L. and {Costagliola}, F. and {Eckart}, A. and {Fuente}, A. and {Henkel}, C. and {Labiano}, A. and {Mart{\'\i}n}, S. and {M{\'a}rquez}, I. and {Muller}, S. and {Planesas}, P. and {Ramos Almeida}, C. and {Spaans}, M. and {Tacconi}, L.~J. and {van der Werf}, P.~P.},
        title = "{Molecular line emission in NGC 1068 imaged with ALMA. I. An AGN-driven outflow in the dense molecular gas}",
      journal = {\aap},
         year = 2014,
        month = jul,
       volume = {567},
          eid = {A125},
        pages = {A125},
          doi = {10.1051/0004-6361/201423843},
archivePrefix = {arXiv},
       eprint = {1405.7706},
 primaryClass = {astro-ph.GA},
       adsurl = {https://ui.adsabs.harvard.edu/abs/2014A&A...567A.125G}
}

@ARTICLE{Viti2014,
       author = {{Viti}, S. and {Garc{\'\i}a-Burillo}, S. and {Fuente}, A. and {Hunt}, L.~K. and {Usero}, A. and {Henkel}, C. and {Eckart}, A. and {Martin}, S. and {Spaans}, M. and {Muller}, S. and {Combes}, F. and {Krips}, M. and {Schinnerer}, E. and {Casasola}, V. and {Costagliola}, F. and {Marquez}, I. and {Planesas}, P. and {van der Werf}, P.~P. and {Aalto}, S. and {Baker}, A.~J. and {Boone}, F. and {Tacconi}, L.~J.},
        title = "{Molecular line emission in NGC 1068 imaged with ALMA. II. The chemistry of the dense molecular gas}",
      journal = {\aap},
         year = 2014,
        month = oct,
       volume = {570},
          eid = {A28},
        pages = {A28},
          doi = {10.1051/0004-6361/201424116},
archivePrefix = {arXiv},
       eprint = {1407.4940},
 primaryClass = {astro-ph.GA},
       adsurl = {https://ui.adsabs.harvard.edu/abs/2014A&A...570A..28V}
}

@ARTICLE{Armus2007,
       author = {{Armus}, L. and {Charmandaris}, V. and {Bernard-Salas}, J. and {Spoon}, H.~W.~W. and {Marshall}, J.~A. and {Higdon}, S.~J.~U. and {Desai}, V. and {Teplitz}, H.~I. and {Hao}, L. and {Devost}, D. and {Brandl}, B.~R. and {Wu}, Y. and {Sloan}, G.~C. and {Soifer}, B.~T. and {Houck}, J.~R. and {Herter}, T.~L.},
        title = "{Observations of Ultraluminous Infrared Galaxies with the Infrared Spectrograph on the Spitzer Space Telescope. II. The IRAS Bright Galaxy Sample}",
      journal = {\apj},
         year = 2007,
        month = feb,
       volume = {656},
       number = {1},
        pages = {148-167},
          doi = {10.1086/510107},
archivePrefix = {arXiv},
       eprint = {astro-ph/0610218},
 primaryClass = {astro-ph},
       adsurl = {https://ui.adsabs.harvard.edu/abs/2007ApJ...656..148A}
}

@ARTICLE{Ramos2014,
       author = {{Ramos Almeida}, C. and {Alonso-Herrero}, A. and {Levenson}, N.~A. and {Asensio Ramos}, A. and {Rodr{\'\i}guez Espinosa}, J.~M. and {Gonz{\'a}lez-Mart{\'\i}n}, O. and {Packham}, C. and {Mart{\'\i}nez}, M.},
        title = "{Investigating the sensitivity of observed spectral energy distributions to clumpy torus properties in Seyfert galaxies}",
      journal = {\mnras},
         year = 2014,
        month = apr,
       volume = {439},
       number = {4},
        pages = {3847-3859},
          doi = {10.1093/mnras/stu235},
archivePrefix = {arXiv},
       eprint = {1402.0345},
 primaryClass = {astro-ph.GA},
       adsurl = {https://ui.adsabs.harvard.edu/abs/2014MNRAS.439.3847R}
}

@ARTICLE{RamosAlmeida2023,
       author = {{Ramos Almeida}, C. and {Esparza-Arredondo}, D. and {Gonz{\'a}lez-Mart{\'\i}n}, O. and {Garc{\'\i}a-Bernete}, I. and {Pereira-Santaella}, M. and {Alonso-Herrero}, A. and {Acosta-Pulido}, J.~A. and {Bessiere}, P.~S. and {Levenson}, N.~A. and {Tadhunter}, C.~N. and {Rigopoulou}, D. and {Mart{\'\i}nez-Paredes}, M. and {Cazzoli}, S. and {Garc{\'\i}a-Lorenzo}, B.},
        title = "{Absence of nuclear polycyclic aromatic hydrocarbon emission from a compact starburst: The case of the type-2 quasar Mrk 477}",
      journal = {\aap},
         year = 2023,
        month = jan,
       volume = {669},
          eid = {L5},
        pages = {L5},
          doi = {10.1051/0004-6361/202245409},
archivePrefix = {arXiv},
       eprint = {2212.01258},
 primaryClass = {astro-ph.GA},
       adsurl = {https://ui.adsabs.harvard.edu/abs/2023A&A...669L...5R}
}

@ARTICLE{Harrison2024,
       author = {{Harrison}, Chris M. and {Ramos Almeida}, Cristina},
        title = "{Observational Tests of Active Galactic Nuclei Feedback: An Overview of Approaches and Interpretation}",
      journal = {Galaxies},
         year = 2024,
        month = apr,
       volume = {12},
       number = {2},
          eid = {17},
        pages = {17},
          doi = {10.3390/galaxies12020017},
archivePrefix = {arXiv},
       eprint = {2404.08050},
 primaryClass = {astro-ph.GA},
       adsurl = {https://ui.adsabs.harvard.edu/abs/2024Galax..12...17H}
}

@ARTICLE{Chen2025,
       author = {{Chen}, Kejian and {Li}, Zhengrong and {Inayoshi}, Kohei and {Ho}, Luis C.},
        title = "{Dust Budget Crisis in Little Red Dots}",
      journal = {\apjl},
         year = 2025,
        month = dec,
       volume = {994},
       number = {2},
          eid = {L42},
        pages = {L42},
          doi = {10.3847/2041-8213/ae1955},
archivePrefix = {arXiv},
       eprint = {2505.22600},
 primaryClass = {astro-ph.GA},
       adsurl = {https://ui.adsabs.harvard.edu/abs/2025ApJ...994L..42C}
}

@software{Diaz-Santos2025,
       author = {{Diaz-Santos}, Tanio and {Lai}, Thomas S.-Y. and {Finnerty}, Luke and {Privon}, George and {Bonfini}, Paolo and {Larson}, Kirsten and {Marshall}, Jason and {Armus}, Lee and {Charmandaris}, Vassilis},
        title = "{CAFE: Continuum And Feature Extraction tool}",
 howpublished = {Astrophysics Source Code Library, record ascl:2501.001},
         year = 2025,
        month = jan,
          eid = {ascl:2501.001},
archivePrefix = {ascl},
       eprint = {2501.001},
       adsurl = {https://ui.adsabs.harvard.edu/abs/2025ascl.soft01001D}
}

@ARTICLE{Donnelly2024,
       author = {{Donnelly}, G.~P. and {Smith}, J.~D.~T. and {Draine}, B.~T. and {Togi}, A. and {Lai}, T.~S.-Y. and {Armus}, L. and {Dale}, D.~A. and {Charmandaris}, V.},
        title = "{The Impact of an Active Galactic Nucleus on Polycyclic Aromatic Hydrocarbon Emission in Galaxies: The Case of Ring Galaxy NGC 4138}",
      journal = {\apj},
         year = 2024,
        month = apr,
       volume = {965},
       number = {1},
          eid = {75},
        pages = {75},
          doi = {10.3847/1538-4357/ad2169},
archivePrefix = {arXiv},
       eprint = {2402.08123},
 primaryClass = {astro-ph.GA},
       adsurl = {https://ui.adsabs.harvard.edu/abs/2024ApJ...965...75D}
}

@ARTICLE{Isbell2022,
       author = {{Isbell}, J.~W. and {Meisenheimer}, K. and {Pott}, J.-U. and {Stalevski}, M. and {Tristram}, K.~R.~W. and {Sanchez-Bermudez}, J. and {Hofmann}, K.-H. and {G{\'a}mez Rosas}, V. and {Jaffe}, W. and {Burtscher}, L. and {Leftley}, J. and {Petrov}, R. and {Lopez}, B. and {Henning}, T. and {Weigelt}, G. and {Allouche}, F. and {Berio}, P. and {Bettonvil}, F. and {Cruzalebes}, P. and {Dominik}, C. and {Heininger}, M. and {Hogerheijde}, M. and {Lagarde}, S. and {Lehmitz}, M. and {Matter}, A. and {Meilland}, A. and {Millour}, F. and {Robbe-Dubois}, S. and {Schertl}, D. and {van Boekel}, R. and {Varga}, J. and {Woillez}, J.},
        title = "{The dusty heart of Circinus. I. Imaging the circumnuclear dust in N-band}",
      journal = {\aap},
         year = 2022,
        month = jul,
       volume = {663},
          eid = {A35},
        pages = {A35},
          doi = {10.1051/0004-6361/202243271},
archivePrefix = {arXiv},
       eprint = {2205.01575},
 primaryClass = {astro-ph.GA},
       adsurl = {https://ui.adsabs.harvard.edu/abs/2022A&A...663A..35I}
}

@ARTICLE{GamezRosas2022,
       author = {{G{\'a}mez Rosas}, Violeta and {Isbell}, Jacob W. and {Jaffe}, Walter and {Petrov}, Romain G. and {Leftley}, James H. and {Hofmann}, Karl-Heinz and {Millour}, Florentin and {Burtscher}, Leonard and {Meisenheimer}, Klaus and {Meilland}, Anthony and {Waters}, Laurens B.~F.~M. and {Lopez}, Bruno and {Lagarde}, St{\'e}phane and {Weigelt}, Gerd and {Berio}, Philippe and {Allouche}, Fatme and {Robbe-Dubois}, Sylvie and {Cruzal{\`e}bes}, Pierre and {Bettonvil}, Felix and {Henning}, Thomas and {Augereau}, Jean-Charles and {Antonelli}, Pierre and {Beckmann}, Udo and {van Boekel}, Roy and {Bendjoya}, Philippe and {Danchi}, William C. and {Dominik}, Carsten and {Drevon}, Julien and {Gallimore}, Jack F. and {Graser}, Uwe and {Heininger}, Matthias and {Hocd{\'e}}, Vincent and {Hogerheijde}, Michiel and {Hron}, Josef and {Impellizzeri}, Caterina M.~V. and {Klarmann}, Lucia and {Kokoulina}, Elena and {Labadie}, Lucas and {Lehmitz}, Michael and {Matter}, Alexis and {Paladini}, Claudia and {Pantin}, Eric and {Pott}, J{\"o}rg-Uwe and {Schertl}, Dieter and {Soulain}, Anthony and {Stee}, Philippe and {Tristram}, Konrad and {Varga}, Jozsef and {Woillez}, Julien and {Wolf}, Sebastian and {Yoffe}, Gideon and {Zins}, Gerard},
        title = "{Thermal imaging of dust hiding the black hole in NGC 1068}",
      journal = {\nat},
         year = 2022,
        month = feb,
       volume = {602},
       number = {7897},
        pages = {403-407},
          doi = {10.1038/s41586-021-04311-7},
archivePrefix = {arXiv},
       eprint = {2112.13694},
 primaryClass = {astro-ph.GA},
       adsurl = {https://ui.adsabs.harvard.edu/abs/2022Natur.602..403G}
}

@ARTICLE{Asmus2019,
       author = {{Asmus}, D.},
        title = "{New evidence for the ubiquity of prominent polar dust emission in AGN on tens of parsec scales}",
      journal = {\mnras},
         year = 2019,
        month = oct,
       volume = {489},
       number = {2},
        pages = {2177-2188},
          doi = {10.1093/mnras/stz2289},
archivePrefix = {arXiv},
       eprint = {1908.03552},
 primaryClass = {astro-ph.GA},
       adsurl = {https://ui.adsabs.harvard.edu/abs/2019MNRAS.489.2177A}
}

@ARTICLE{Stalevski2017,
       author = {{Stalevski}, Marko and {Asmus}, Daniel and {Tristram}, Konrad R.~W.},
        title = "{Dissecting the active galactic nucleus in Circinus - I. Peculiar mid-IR morphology explained by a dusty hollow cone}",
      journal = {\mnras},
         year = 2017,
        month = dec,
       volume = {472},
       number = {4},
        pages = {3854-3870},
          doi = {10.1093/mnras/stx2227},
archivePrefix = {arXiv},
       eprint = {1708.07838},
 primaryClass = {astro-ph.GA},
       adsurl = {https://ui.adsabs.harvard.edu/abs/2017MNRAS.472.3854S}
}

@ARTICLE{Haidar2026,
       author = {{Haidar}, Houda and {Rosario}, David J. and {Garc{\'\i}a-Bernete}, Ismael and {Alonso-Herrero}, Almudena and {Audibert}, Anelise and {Campbell}, Steph and {Harrison}, Chris M. and {Costa}, Tiago and {Hermosa Mu{\~n}oz}, Laura and {Combes}, Fran{\c{c}}oise and {Rigopoulou}, Dimitra and {Ricci}, Claudio and {Ramos Almeida}, Cristina and {Bellocchi}, Enrica and {Boorman}, Peter and {Bunker}, Andrew and {Davies}, Richard and {Delaney}, Daniel and {D{\'\i}az Santos}, Tanio and {Esposito}, Federico and {Fawcett}, Victoria A. and {Gandhi}, Poshak and {Garc{\'\i}a-Burillo}, Santiago and {Gonz{\'a}lez-Mart{\'\i}n}, Omaira and {Hicks}, Erin K.~S. and {H{\"o}nig}, Sebastian F. and {Labiano}, Alvaro and {Levenson}, Nancy A. and {Lopez-Rodriguez}, Enrique and {Packham}, Chris and {Pereira-Santaella}, Miguel and {Riffel}, Rogemar A. and {Rodr{\'\i}guez Ardila}, Alberto and {Schneider}, John and {Shimizu}, T. Taro and {Stalevski}, Marko and {Villar Mart{\'\i}n}, Montserrat and {Ward}, Martin and {Zhang}, Lulu and {Leeds}, Gillian and {Donnan}, Fergus R.},
        title = "{GATOS XI : Excess dust heating in the Narrow Line Regions of nearby AGN revealed with JWST/MIRI}",
      journal = {arXiv e-prints},
         year = 2026,
        month = jan,
          eid = {arXiv:2601.02865},
        pages = {arXiv:2601.02865},
          doi = {10.48550/arXiv.2601.02865},
archivePrefix = {arXiv},
       eprint = {2601.02865},
 primaryClass = {astro-ph.GA},
       adsurl = {https://ui.adsabs.harvard.edu/abs/2026arXiv260102865H}
}

@ARTICLE{Smith2007,
       author = {{Smith}, J.~D.~T. and {Draine}, B.~T. and {Dale}, D.~A. and {Moustakas}, J. and {Kennicutt}, R.~C., Jr. and {Helou}, G. and {Armus}, L. and {Roussel}, H. and {Sheth}, K. and {Bendo}, G.~J. and {Buckalew}, B.~A. and {Calzetti}, D. and {Engelbracht}, C.~W. and {Gordon}, K.~D. and {Hollenbach}, D.~J. and {Li}, A. and {Malhotra}, S. and {Murphy}, E.~J. and {Walter}, F.},
        title = "{The Mid-Infrared Spectrum of Star-forming Galaxies: Global Properties of Polycyclic Aromatic Hydrocarbon Emission}",
      journal = {\apj},
         year = 2007,
        month = feb,
       volume = {656},
       number = {2},
        pages = {770-791},
          doi = {10.1086/510549},
archivePrefix = {arXiv},
       eprint = {astro-ph/0610913},
 primaryClass = {astro-ph},
       adsurl = {https://ui.adsabs.harvard.edu/abs/2007ApJ...656..770S}
}

@ARTICLE{Canelo2026,
       author = {{Canelo}, Carla M. and {Sales}, Dinalva A. and {Avelaneda}, Vitor N. and {Tielens}, Alexander G.~G.~M. and {Pastoriza}, Miriani G. and {Fria{\c{c}}a}, Am{\^a}ncio C.~S.},
        title = "{Variations in the 6.2 {\ensuremath{\mu}}m Polycyclic Aromatic Hydrocarbon Band in Active Galactic Nucleus- and Starburst-dominated Galaxies}",
      journal = {\apj},
         year = 2026,
        month = apr,
       volume = {1000},
       number = {2},
          eid = {171},
        pages = {171},
          doi = {10.3847/1538-4357/ae45a5},
archivePrefix = {arXiv},
       eprint = {2603.18901},
 primaryClass = {astro-ph.GA},
       adsurl = {https://ui.adsabs.harvard.edu/abs/2026ApJ..1000..171C}
}

@ARTICLE{Fabian2006,
       author = {{Fabian}, A.~C. and {Celotti}, A. and {Erlund}, M.~C.},
        title = "{Radiative pressure feedback by a quasar in a galactic bulge}",
      journal = {\mnras},
         year = 2006,
        month = nov,
       volume = {373},
       number = {1},
        pages = {L16-L20},
          doi = {10.1111/j.1745-3933.2006.00234.x},
archivePrefix = {arXiv},
       eprint = {astro-ph/0608425},
 primaryClass = {astro-ph},
       adsurl = {https://ui.adsabs.harvard.edu/abs/2006MNRAS.373L..16F}
}

@ARTICLE{Zhang2024b,
       author = {{Zhang}, Lulu and {Packham}, Chris and {Hicks}, Erin K.~S. and {Davies}, Ric I. and {Shimizu}, Taro T. and {Alonso-Herrero}, Almudena and {Hermosa Mu{\~n}oz}, Laura and {Garc{\'\i}a-Bernete}, Ismael and {Pereira-Santaella}, Miguel and {Audibert}, Anelise and {L{\'o}pez-Rodr{\'\i}guez}, Enrique and {Bellocchi}, Enrica and {Bunker}, Andrew J. and {Combes}, Francoise and {D{\'\i}az-Santos}, Tanio and {Gandhi}, Poshak and {Garc{\'\i}a-Burillo}, Santiago and {Garc{\'\i}a-Lorenzo}, Bego{\~n}a and {Gonz{\'a}lez-Mart{\'\i}n}, Omaira and {Imanishi}, Masatoshi and {Labiano}, Alvaro and {Leist}, Mason T. and {Levenson}, Nancy A. and {Ramos Almeida}, Cristina and {Ricci}, Claudio and {Rigopoulou}, Dimitra and {Rosario}, David J. and {Stalevski}, Marko and {Ward}, Martin J. and {Esparza-Arredondo}, Donaji and {Delaney}, Dan and {Fuller}, Lindsay and {Haidar}, Houda and {H{\"o}nig}, Sebastian and {Izumi}, Takuma and {Rouan}, Daniel},
        title = "{The Galaxy Activity, Torus, and Outflow Survey (GATOS). IV. Exploring Ionized Gas Outflows in Central Kiloparsec Regions of GATOS Seyferts}",
      journal = {\apj},
         year = 2024,
        month = oct,
       volume = {974},
       number = {2},
          eid = {195},
        pages = {195},
          doi = {10.3847/1538-4357/ad6a4b},
archivePrefix = {arXiv},
       eprint = {2409.09771},
 primaryClass = {astro-ph.GA},
       adsurl = {https://ui.adsabs.harvard.edu/abs/2024ApJ...974..195Z}
}

@ARTICLE{Davies2024,
       author = {{Davies}, R. and {Shimizu}, T. and {Pereira-Santaella}, M. and {Alonso-Herrero}, A. and {Audibert}, A. and {Bellocchi}, E. and {Boorman}, P. and {Campbell}, S. and {Cao}, Y. and {Combes}, F. and {Delaney}, D. and {D{\'\i}az-Santos}, T. and {Eisenhauer}, F. and {Esparza Arredondo}, D. and {Feuchtgruber}, H. and {F{\"o}rster Schreiber}, N.~M. and {Fuller}, L. and {Gandhi}, P. and {Garc{\'\i}a-Bernete}, I. and {Garc{\'\i}a-Burillo}, S. and {Garc{\'\i}a-Lorenzo}, B. and {Genzel}, R. and {Gillessen}, S. and {Gonz{\'a}lez Mart{\'\i}n}, O. and {Haidar}, H. and {Hermosa Mu{\~n}oz}, L. and {Hicks}, E.~K.~S. and {H{\"o}nig}, S. and {Imanishi}, M. and {Izumi}, T. and {Labiano}, A. and {Leist}, M. and {Levenson}, N.~A. and {Lopez-Rodriguez}, E. and {Lutz}, D. and {Ott}, T. and {Packham}, C. and {Rabien}, S. and {Ramos Almeida}, C. and {Ricci}, C. and {Rigopoulou}, D. and {Rosario}, D. and {Rouan}, D. and {Santos}, D.~J.~D. and {Shangguan}, J. and {Stalevski}, M. and {Sternberg}, A. and {Sturm}, E. and {Tacconi}, L. and {Villar Mart{\'\i}n}, M. and {Ward}, M. and {Zhang}, L.},
        title = "{GATOS: missing molecular gas in the outflow of NGC 5728 revealed by JWST}",
      journal = {\aap},
         year = 2024,
        month = sep,
       volume = {689},
          eid = {A263},
        pages = {A263},
          doi = {10.1051/0004-6361/202449875},
archivePrefix = {arXiv},
       eprint = {2406.17072},
 primaryClass = {astro-ph.GA},
       adsurl = {https://ui.adsabs.harvard.edu/abs/2024A&A...689A.263D}
}

@ARTICLE{Braatz1993,
       author = {{Braatz}, J.~A. and {Wilson}, A.~S. and {Gezari}, D.~Y. and {Varosi}, F. and {Beichman}, C.~A.},
        title = "{High-Resolution Mid-Infrared Imaging and Astrometry of the Nucleus of the Seyfert Galaxy NGC 1068}",
      journal = {\apjl},
         year = 1993,
        month = may,
       volume = {409},
        pages = {L5},
          doi = {10.1086/186846},
       adsurl = {https://ui.adsabs.harvard.edu/abs/1993ApJ...409L...5B}
}

@ARTICLE{Bock2000,
       author = {{Bock}, J.~J. and {Neugebauer}, G. and {Matthews}, K. and {Soifer}, B.~T. and {Becklin}, E.~E. and {Ressler}, M. and {Marsh}, K. and {Werner}, M.~W. and {Egami}, E. and {Blandford}, R.},
        title = "{High Spatial Resolution Imaging of NGC 1068 in the Mid-Infrared}",
      journal = {\aj},
         year = 2000,
        month = dec,
       volume = {120},
       number = {6},
        pages = {2904-2919},
          doi = {10.1086/316871},
archivePrefix = {arXiv},
       eprint = {astro-ph/0009078},
 primaryClass = {astro-ph},
       adsurl = {https://ui.adsabs.harvard.edu/abs/2000AJ....120.2904B}
}

@ARTICLE{Kang2018,
       author = {{Kang}, Daeun and {Woo}, Jong-Hak},
        title = "{Unraveling the Complex Structure of AGN-driven Outflows. III. The Outflow Size-Luminosity Relation}",
      journal = {\apj},
         year = 2018,
        month = sep,
       volume = {864},
       number = {2},
          eid = {124},
        pages = {124},
          doi = {10.3847/1538-4357/aad561},
archivePrefix = {arXiv},
       eprint = {1807.08356},
 primaryClass = {astro-ph.GA},
       adsurl = {https://ui.adsabs.harvard.edu/abs/2018ApJ...864..124K}
}

@ARTICLE{Kim2023,
       author = {{Kim}, Changseok and {Woo}, Jong-Hak and {Luo}, Rongxin and {Chung}, Aeree and {Baek}, Junhyun and {Le}, Huynh Anh N. and {Son}, Donghoon},
        title = "{Unraveling the Complex Structure of AGN-driven Outflows. VI. Strong Ionized Outflows in Type 1 AGNs and the Outflow Size-Luminosity Relation}",
      journal = {\apj},
         year = 2023,
        month = dec,
       volume = {958},
       number = {2},
          eid = {145},
        pages = {145},
          doi = {10.3847/1538-4357/acf92b},
archivePrefix = {arXiv},
       eprint = {2310.06928},
 primaryClass = {astro-ph.GA},
       adsurl = {https://ui.adsabs.harvard.edu/abs/2023ApJ...958..145K}
}

@ARTICLE{Kakkad2022,
       author = {{Kakkad}, D. and {Sani}, E. and {Rojas}, A.~F. and {Mallmann}, Nicolas D. and {Veilleux}, S. and {Bauer}, Franz E. and {Ricci}, F. and {Mushotzky}, R. and {Koss}, M. and {Ricci}, C. and {Treister}, E. and {Privon}, George C. and {Nguyen}, N. and {B{\"a}r}, R. and {Harrison}, F. and {Oh}, K. and {Powell}, M. and {Riffel}, R. and {Stern}, D. and {Trakhtenbrot}, B. and {Urry}, C.~M.},
        title = "{BASS XXXI: Outflow scaling relations in low redshift X-ray AGN host galaxies with MUSE}",
      journal = {\mnras},
         year = 2022,
        month = apr,
       volume = {511},
       number = {2},
        pages = {2105-2124},
          doi = {10.1093/mnras/stac103},
archivePrefix = {arXiv},
       eprint = {2201.04149},
 primaryClass = {astro-ph.GA},
       adsurl = {https://ui.adsabs.harvard.edu/abs/2022MNRAS.511.2105K}
}

@ARTICLE{Zhang2026,
       author = {{Zhang}, Lulu and {Packham}, Chris and {Hicks}, Erin K.~S. and {Davies}, Ric I. and {Delaney}, Daniel E. and {Combes}, Francoise and {Pereira-Santaella}, Miguel and {Alonso-Herrero}, Almudena and {Ricci}, Claudio and {Gonz{\'a}lez-Mart{\'\i}n}, Omaira and {Hermosa Mu{\~n}oz}, Laura and {Garc{\'\i}a-Bernete}, Ismael and {Ramos Almeida}, Cristina and {Rigopoulou}, Dimitra and {Donnan}, Fergus R. and {Bellocchi}, Enrica and {Levenson}, Nancy A. and {Ward}, Martin J. and {Garc{\'\i}a-Burillo}, Santiago and {Hoenig}, Sebastian F.},
        title = "{Evidence of Feedback Effects in Low-luminosity Active Galactic Nuclei Revealed by JWST Spectroscopy}",
      journal = {\apjl},
         year = 2026,
        month = feb,
       volume = {998},
       number = {2},
          eid = {L32},
        pages = {L32},
          doi = {10.3847/2041-8213/ae3f32},
archivePrefix = {arXiv},
       eprint = {2601.07990},
 primaryClass = {astro-ph.GA},
       adsurl = {https://ui.adsabs.harvard.edu/abs/2026ApJ...998L..32Z}
}

@ARTICLE{Radomski2002,
       author = {{Radomski}, James T. and {Pi{\~n}a}, Robert K. and {Packham}, Christopher and {Telesco}, Charles M. and {Tadhunter}, Clive N.},
        title = "{High-Resolution Mid-Infrared Morphology of Cygnus A}",
      journal = {\apj},
         year = 2002,
        month = feb,
       volume = {566},
       number = {2},
        pages = {675-681},
          doi = {10.1086/338071},
archivePrefix = {arXiv},
       eprint = {astro-ph/0110347},
 primaryClass = {astro-ph},
       adsurl = {https://ui.adsabs.harvard.edu/abs/2002ApJ...566..675R}
}

@ARTICLE{Packham2005,
       author = {{Packham}, Christopher and {Radomski}, James T. and {Roche}, Patrick F. and {Aitken}, David K. and {Perlman}, Eric and {Alonso-Herrero}, Almudena and {Colina}, Luis and {Telesco}, Charles M.},
        title = "{The Extended Mid-Infrared Structure of the Circinus Galaxy}",
      journal = {\apjl},
         year = 2005,
        month = jan,
       volume = {618},
       number = {1},
        pages = {L17-L20},
          doi = {10.1086/427691},
       adsurl = {https://ui.adsabs.harvard.edu/abs/2005ApJ...618L..17P}
}

@ARTICLE{Radomski2008,
       author = {{Radomski}, James T. and {Packham}, Christopher and {Levenson}, N.~A. and {Perlman}, Eric and {Leeuw}, Lerothodi L. and {Matthews}, Henry and {Mason}, Rachel and {De Buizer}, James M. and {Telesco}, Charles M. and {Orduna}, Manuel},
        title = "{Gemini Imaging of Mid-Infrared Emission from the Nuclear Region of Centaurus A}",
      journal = {\apj},
         year = 2008,
        month = jul,
       volume = {681},
       number = {1},
        pages = {141-150},
          doi = {10.1086/587771},
archivePrefix = {arXiv},
       eprint = {0802.4119},
 primaryClass = {astro-ph},
       adsurl = {https://ui.adsabs.harvard.edu/abs/2008ApJ...681..141R}
}

@ARTICLE{Reunanen2010,
       author = {{Reunanen}, J. and {Prieto}, M.~A. and {Siebenmorgen}, R.},
        title = "{VLT diffraction-limited imaging at 11 and 18{\ensuremath{\mu}}m of the nearest active galactic nuclei}",
      journal = {\mnras},
         year = 2010,
        month = feb,
       volume = {402},
       number = {2},
        pages = {879-894},
          doi = {10.1111/j.1365-2966.2009.15997.x},
archivePrefix = {arXiv},
       eprint = {0911.2112},
 primaryClass = {astro-ph.CO},
       adsurl = {https://ui.adsabs.harvard.edu/abs/2010MNRAS.402..879R}
}

@ARTICLE{Asmus2014,
       author = {{Asmus}, D. and {H{\"o}nig}, S.~F. and {Gandhi}, P. and {Smette}, A. and {Duschl}, W.~J.},
        title = "{The subarcsecond mid-infrared view of local active galactic nuclei - I. The N- and Q-band imaging atlas}",
      journal = {\mnras},
         year = 2014,
        month = apr,
       volume = {439},
       number = {2},
        pages = {1648-1679},
          doi = {10.1093/mnras/stu041},
archivePrefix = {arXiv},
       eprint = {1310.2770},
 primaryClass = {astro-ph.CO},
       adsurl = {https://ui.adsabs.harvard.edu/abs/2014MNRAS.439.1648A}
}

@BOOK{Tielens2010,
       author = {{Tielens}, A.~G.~G.~M.},
        title = "{The Physics and Chemistry of the Interstellar Medium}",
         year = 2010,
       adsurl = {https://ui.adsabs.harvard.edu/abs/2010pcim.book.....T}
}

@ARTICLE{Garcia-Bernete2016,
       author = {{Garc{\'\i}a-Bernete}, I. and {Ramos Almeida}, C. and {Acosta-Pulido}, J.~A. and {Alonso-Herrero}, A. and {Gonz{\'a}lez-Mart{\'\i}n}, O. and {Hern{\'a}n-Caballero}, A. and {Pereira-Santaella}, M. and {Levenson}, N.~A. and {Packham}, C. and {Perlman}, E.~S. and {Ichikawa}, K. and {Esquej}, P. and {D{\'\i}az-Santos}, T.},
        title = "{The nuclear and extended mid-infrared emission of Seyfert galaxies}",
      journal = {\mnras},
         year = 2016,
        month = dec,
       volume = {463},
       number = {4},
        pages = {3531-3555},
          doi = {10.1093/mnras/stw2125},
archivePrefix = {arXiv},
       eprint = {1608.06513},
 primaryClass = {astro-ph.GA},
       adsurl = {https://ui.adsabs.harvard.edu/abs/2016MNRAS.463.3531G}
}

@ARTICLE{Donnan24b,
       author = {{Donnan}, Fergus R and {Rigopoulou}, Dimitra and {Garc{\'\i}a-Bernete}, Ismael},
        title = "{The kinematics of Polycyclic Aromatic Hydrocarbons (PAHs) in Galaxies revealed by Principal Component Analysis (PCA) tomography with JWST/NIRSpec}",
      journal = {\mnras},
         year = 2024,
        month = jul,
       volume = {532},
       number = {1},
        pages = {L75-L81},
          doi = {10.1093/mnrasl/slae050},
archivePrefix = {arXiv},
       eprint = {2405.13669},
 primaryClass = {astro-ph.GA},
       adsurl = {https://ui.adsabs.harvard.edu/abs/2024MNRAS.532L..75D}
}

@ARTICLE{Fabian2008,
       author = {{Fabian}, A.~C. and {Vasudevan}, R.~V. and {Gandhi}, P.},
        title = "{The effect of radiation pressure on dusty absorbing gas around active galactic nuclei}",
      journal = {\mnras},
         year = 2008,
        month = mar,
       volume = {385},
       number = {1},
        pages = {L43-L47},
          doi = {10.1111/j.1745-3933.2008.00430.x},
archivePrefix = {arXiv},
       eprint = {0712.0277},
 primaryClass = {astro-ph},
       adsurl = {https://ui.adsabs.harvard.edu/abs/2008MNRAS.385L..43F}
}

@ARTICLE{Venanzi2020,
       author = {{Venanzi}, Marta and {H{\"o}nig}, Sebastian and {Williamson}, David},
        title = "{The Role of Infrared Radiation Pressure in Shaping Dusty Winds in AGNs}",
      journal = {\apj},
         year = 2020,
        month = sep,
       volume = {900},
       number = {2},
          eid = {174},
        pages = {174},
          doi = {10.3847/1538-4357/aba89f},
archivePrefix = {arXiv},
       eprint = {2007.13554},
 primaryClass = {astro-ph.GA},
       adsurl = {https://ui.adsabs.harvard.edu/abs/2020ApJ...900..174V}
}

@ARTICLE{Campbell2025,
       author = {{Campbell}, Steph and {Rosario}, David J. and {Haidar}, Houda and {L{\'o}pez Rodr{\'\i}guez}, Enrique and {Delaney}, Dan and {Hicks}, Erin and {Garc{\'\i}a-Bernete}, Ismael and {Pereira-Santaella}, Miguel and {Alonso Herrero}, Almudena and {Audibert}, Anelise and {Bellocchi}, Enrica and {Esparza-Arredondo}, Donaji and {Garc{\'\i}a-Burillo}, Santiago and {Gonz{\'a}lez Mart{\'\i}n}, Omaira and {H{\"o}nig}, Sebastian F. and {Levenson}, Nancy A. and {Packham}, Chris and {Ramos Almeida}, Cristina and {Rigopoulou}, Dimitra and {Zhang}, Lulu},
        title = "{GATOS ─ IX. A detailed assessment and treatment of emission line contamination in JWST/MIRI images of nearby Seyfert galaxies}",
      journal = {\mnras},
         year = 2025,
        month = nov,
       volume = {544},
       number = {1},
        pages = {648-668},
          doi = {10.1093/mnras/staf1719},
archivePrefix = {arXiv},
       eprint = {2510.09830},
 primaryClass = {astro-ph.GA},
       adsurl = {https://ui.adsabs.harvard.edu/abs/2025MNRAS.544..648C}
}

@ARTICLE{Garcia-Bernete2024b,
       author = {{Garc{\'\i}a-Bernete}, I. and {Rigopoulou}, D. and {Donnan}, F.~R. and {Alonso-Herrero}, A. and {Pereira-Santaella}, M. and {Shimizu}, T. and {Davies}, R. and {Roche}, P.~F. and {Garc{\'\i}a-Burillo}, S. and {Labiano}, A. and {Hermosa Mu{\~n}oz}, L. and {Zhang}, L. and {Audibert}, A. and {Bellocchi}, E. and {Bunker}, A. and {Combes}, F. and {Delaney}, D. and {Esparza-Arredondo}, D. and {Gandhi}, P. and {Gonz{\'a}lez-Mart{\'\i}n}, O. and {H{\"o}nig}, S.~F. and {Imanishi}, M. and {Hicks}, E.~K.~S. and {Fuller}, L. and {Leist}, M. and {Levenson}, N.~A. and {Lopez-Rodriguez}, E. and {Packham}, C. and {Ramos Almeida}, C. and {Ricci}, C. and {Stalevski}, M. and {Villar Mart{\'\i}n}, M. and {Ward}, M.~J.},
        title = "{The Galaxy Activity, Torus, and Outflow Survey (GATOS): V. Unveiling PAH survival and resilience in the circumnuclear regions of AGNs with JWST}",
      journal = {\aap},
         year = 2024,
        month = nov,
       volume = {691},
          eid = {A162},
        pages = {A162},
          doi = {10.1051/0004-6361/202450086},
archivePrefix = {arXiv},
       eprint = {2409.05686},
 primaryClass = {astro-ph.GA},
       adsurl = {https://ui.adsabs.harvard.edu/abs/2024A&A...691A.162G}
}

@ARTICLE{Rigopoulou2024,
       author = {{Rigopoulou}, D. and {Donnan}, F.~R. and {Garc{\'\i}a-Bernete}, I. and {Pereira-Santaella}, M. and {Alonso-Herrero}, A. and {Davies}, R. and {Hunt}, L.~K. and {Roche}, P.~F. and {Shimizu}, T.},
        title = "{Polycyclic aromatic hydrocarbon emission in galaxies as seen with JWST}",
      journal = {\mnras},
         year = 2024,
        month = aug,
       volume = {532},
       number = {2},
        pages = {1598-1611},
          doi = {10.1093/mnras/stae1535},
archivePrefix = {arXiv},
       eprint = {2406.11415},
 primaryClass = {astro-ph.GA},
       adsurl = {https://ui.adsabs.harvard.edu/abs/2024MNRAS.532.1598R}
}

@ARTICLE{Veenema2025,
       author = {{Veenema}, Oscar and {Thatte}, Niranjan and {Rigopoulou}, Dimitra and {Garc{\'\i}a-Bernete}, Ismael and {Alonso-Herrero}, Almudena and {Audibert}, Anelise and {Bellocchi}, Enrica and {Bunker}, Andrew J. and {Campbell}, Steph and {Combes}, Francoise and {Davies}, Ric I. and {Delaney}, Daniel and {Donnan}, Fergus and {Esposito}, Federico and {Garc{\'\i}a-Burillo}, Santiago and {Gonzalez Martin}, Omaira and {Hermosa Mu{\~n}oz}, Laura and {Hicks}, Erin K.~S. and {Hoenig}, Sebastian F. and {Levenson}, Nancy A. and {Packham}, Chris and {Pereira-Santaella}, Miguel and {Ramos Almeida}, Cristina and {Ricci}, Claudio and {Riffel}, Rogemar A. and {Rosario}, David and {Zhang}, Lulu},
        title = "{Shock-driven heating in the circumnuclear star-forming regions of NGC 7582: insights from JWST NIRSpec and MIRI/MRS spectroscopy}",
      journal = {\mnras},
         year = 2025,
        month = dec,
       volume = {544},
       number = {4},
        pages = {3361-3378},
          doi = {10.1093/mnras/staf1887},
archivePrefix = {arXiv},
       eprint = {2510.26321},
 primaryClass = {astro-ph.GA},
       adsurl = {https://ui.adsabs.harvard.edu/abs/2025MNRAS.544.3361V}
}

@ARTICLE{RamosAlmeida2017,
       author = {{Ramos Almeida}, Cristina and {Ricci}, Claudio},
        title = "{Nuclear obscuration in active galactic nuclei}",
      journal = {Nature Astronomy},
         year = 2017,
        month = oct,
       volume = {1},
        pages = {679-689},
          doi = {10.1038/s41550-017-0232-z},
archivePrefix = {arXiv},
       eprint = {1709.00019},
 primaryClass = {astro-ph.GA},
       adsurl = {https://ui.adsabs.harvard.edu/abs/2017NatAs...1..679R}
}

@ARTICLE{Audibert25,
       author = {{Audibert}, A. and {Ramos Almeida}, C. and {Garc{\'\i}a-Burillo}, S. and {Speranza}, G. and {Lamperti}, I. and {Pereira-Santaella}, M. and {Panessa}, F.},
        title = "{Molecular gas excitation and outflow properties of obscured quasars at z {\ensuremath{\sim}} 0.1}",
      journal = {\aap},
         year = 2025,
        month = jul,
       volume = {699},
          eid = {A83},
        pages = {A83},
          doi = {10.1051/0004-6361/202453291},
archivePrefix = {arXiv},
       eprint = {2505.02759},
 primaryClass = {astro-ph.GA},
       adsurl = {https://ui.adsabs.harvard.edu/abs/2025A&A...699A..83A}
}

@ARTICLE{Speranza24,
       author = {{Speranza}, G. and {Ramos Almeida}, C. and {Acosta-Pulido}, J.~A. and {Audibert}, A. and {Holden}, L.~R. and {Tadhunter}, C.~N. and {Lapi}, A. and {Gonz{\'a}lez-Mart{\'\i}n}, O. and {Brusa}, M. and {L{\'o}pez}, I.~E. and {Musiimenta}, B. and {Shankar}, F.},
        title = "{Multiphase characterization of AGN winds in five local type-2 quasars}",
      journal = {\aap},
         year = 2024,
        month = jan,
       volume = {681},
          eid = {A63},
        pages = {A63},
          doi = {10.1051/0004-6361/202347715},
archivePrefix = {arXiv},
       eprint = {2311.10132},
 primaryClass = {astro-ph.GA},
       adsurl = {https://ui.adsabs.harvard.edu/abs/2024A&A...681A..63S}
}

@ARTICLE{RamosAlmeida2025,
       author = {{Ramos Almeida}, C. and {Garc{\'\i}a-Bernete}, I. and {Pereira-Santaella}, M. and {Speranza}, G. and {Maiolino}, R. and {Ji}, X. and {Audibert}, A. and {Cezar}, P.~H. and {Acosta-Pulido}, J.~A. and {Alonso-Herrero}, A. and {Garc{\'\i}a-Burillo}, S. and {Gonz{\'a}lez-Mart{\'\i}n}, O. and {Rigopoulou}, D. and {Tadhunter}, C.~N. and {Labiano}, A. and {Levenson}, N.~A. and {Donnan}, F.~R.},
        title = "{JWST MIRI reveals the diversity of nuclear mid-infrared spectra of nearby type 2 quasars}",
      journal = {\aap},
         year = 2025,
        month = jun,
       volume = {698},
          eid = {A194},
        pages = {A194},
          doi = {10.1051/0004-6361/202453549},
archivePrefix = {arXiv},
       eprint = {2504.01595},
 primaryClass = {astro-ph.GA},
       adsurl = {https://ui.adsabs.harvard.edu/abs/2025A&A...698A.194R}
}

@ARTICLE{Alonso-Herrero2014,
       author = {{Alonso-Herrero}, A. and {Ramos Almeida}, C. and {Esquej}, P. and {Roche}, P.~F. and {Hern{\'a}n-Caballero}, A. and {H{\"o}nig}, S.~F. and {Gonz{\'a}lez-Mart{\'\i}n}, O. and {Aretxaga}, I. and {Mason}, R.~E. and {Packham}, C. and {Levenson}, N.~A. and {Rodr{\'\i}guez Espinosa}, J.~M. and {Siebenmorgen}, R. and {Pereira-Santaella}, M. and {D{\'\i}az-Santos}, T. and {Colina}, L. and {Alvarez}, C. and {Telesco}, C.~M.},
        title = "{Nuclear 11.3 {\ensuremath{\mu}}m PAH emission in local active galactic nuclei}",
      journal = {\mnras},
         year = 2014,
        month = sep,
       volume = {443},
       number = {3},
        pages = {2766-2782},
          doi = {10.1093/mnras/stu1293},
archivePrefix = {arXiv},
       eprint = {1407.1154},
 primaryClass = {astro-ph.GA},
       adsurl = {https://ui.adsabs.harvard.edu/abs/2014MNRAS.443.2766A}
}

@ARTICLE{HermosaMunoz2024,
       author = {{Hermosa Mu{\~n}oz}, L. and {Alonso-Herrero}, A. and {Pereira-Santaella}, M. and {Garc{\'\i}a-Bernete}, I. and {Garc{\'\i}a-Burillo}, S. and {Garc{\'\i}a-Lorenzo}, B. and {Davies}, R. and {Shimizu}, T. and {Esparza-Arredondo}, D. and {Hicks}, E.~K.~S. and {Haidar}, H. and {Leist}, M. and {L{\'o}pez-Rodr{\'\i}guez}, E. and {Ramos Almeida}, C. and {Rosario}, D. and {Zhang}, L. and {Audibert}, A. and {Bellocchi}, E. and {Boorman}, P. and {Bunker}, A.~J. and {Combes}, F. and {Campbell}, S. and {D{\'\i}az-Santos}, T. and {Fuller}, L. and {Gandhi}, P. and {Gonz{\'a}lez-Mart{\'\i}n}, O. and {H{\"o}nig}, S. and {Imanishi}, M. and {Izumi}, T. and {Labiano}, A. and {Levenson}, N.~A. and {Packham}, C. and {Ricci}, C. and {Rigopoulou}, D. and {Rouan}, D. and {Stalevski}, M. and {Villar-Mart{\'\i}n}, M. and {Ward}, M.~J.},
        title = "{A biconical ionised gas outflow and evidence of positive feedback in NGC 7172 uncovered by MIRI/JWST}",
      journal = {\aap},
         year = 2024,
        month = oct,
       volume = {690},
          eid = {A350},
        pages = {A350},
          doi = {10.1051/0004-6361/202450262},
archivePrefix = {arXiv},
       eprint = {2407.15807},
 primaryClass = {astro-ph.GA},
       adsurl = {https://ui.adsabs.harvard.edu/abs/2024A&A...690A.350H}
}

@ARTICLE{Zhang2024,
       author = {{Zhang}, Lulu and {Garc{\'\i}a-Bernete}, Ismael and {Packham}, Chris and {Donnan}, Fergus R. and {Rigopoulou}, Dimitra and {Hicks}, Erin K.~S. and {Davies}, Ric I. and {Shimizu}, Taro T. and {Alonso-Herrero}, Almudena and {Ramos Almeida}, Cristina and {Pereira-Santaella}, Miguel and {Ricci}, Claudio and {Bunker}, Andrew J. and {Leist}, Mason T. and {Rosario}, David J. and {Garc{\'\i}a-Burillo}, Santiago and {Hermosa Mu{\~n}oz}, Laura and {Combes}, Francoise and {Imanishi}, Masatoshi and {Labiano}, Alvaro and {Esparza-Arredondo}, Donaji and {Bellocchi}, Enrica and {Audibert}, Anelise and {Fuller}, Lindsay and {Gonz{\'a}lez-Mart{\'\i}n}, Omaira and {H{\"o}nig}, Sebastian and {Izumi}, Takuma and {Levenson}, Nancy A. and {L{\'o}pez-Rodr{\'\i}guez}, Enrique and {Rouan}, Daniel and {Stalevski}, Marko and {Ward}, Martin J.},
        title = "{Polycyclic Aromatic Hydrocarbon Emission in the Central Regions of Three Seyferts and the Implication for Underlying Feedback Mechanisms}",
      journal = {\apjl},
         year = 2024,
        month = nov,
       volume = {975},
       number = {1},
          eid = {L2},
        pages = {L2},
          doi = {10.3847/2041-8213/ad81d0},
archivePrefix = {arXiv},
       eprint = {2409.09772},
 primaryClass = {astro-ph.GA},
       adsurl = {https://ui.adsabs.harvard.edu/abs/2024ApJ...975L...2Z}
}

@ARTICLE{Garcia-Burillo2024,
       author = {{Garc{\'\i}a-Burillo}, S. and {Hicks}, E.~K.~S. and {Alonso-Herrero}, A. and {Pereira-Santaella}, M. and {Usero}, A. and {Querejeta}, M. and {Gonz{\'a}lez-Mart{\'\i}n}, O. and {Delaney}, D. and {Ramos Almeida}, C. and {Combes}, F. and {Angl{\'e}s-Alc{\'a}zar}, D. and {Audibert}, A. and {Bellocchi}, E. and {Davies}, R.~I. and {Davis}, T.~A. and {Elford}, J.~S. and {Garc{\'\i}a-Bernete}, I. and {H{\"o}nig}, S. and {Labiano}, A. and {Leist}, M.~T. and {Levenson}, N.~A. and {L{\'o}pez-Rodr{\'\i}guez}, E. and {Mercedes-Feliz}, J. and {Packham}, C. and {Ricci}, C. and {Rosario}, D.~J. and {Shimizu}, T. and {Stalevski}, M. and {Zhang}, L.},
        title = "{Deciphering the imprint of active galactic nucleus feedback in Seyfert galaxies: Nuclear-scale molecular gas deficits}",
      journal = {\aap},
         year = 2024,
        month = sep,
       volume = {689},
          eid = {A347},
        pages = {A347},
          doi = {10.1051/0004-6361/202450268},
archivePrefix = {arXiv},
       eprint = {2406.11398},
 primaryClass = {astro-ph.GA},
       adsurl = {https://ui.adsabs.harvard.edu/abs/2024A&A...689A.347G}
}

@ARTICLE{Ricci2017b,
       author = {{Ricci}, Claudio and {Trakhtenbrot}, Benny and {Koss}, Michael J. and {Ueda}, Yoshihiro and {Schawinski}, Kevin and {Oh}, Kyuseok and {Lamperti}, Isabella and {Mushotzky}, Richard and {Treister}, Ezequiel and {Ho}, Luis C. and {Weigel}, Anna and {Bauer}, Franz E. and {Paltani}, Stephane and {Fabian}, Andrew C. and {Xie}, Yanxia and {Gehrels}, Neil},
        title = "{The close environments of accreting massive black holes are shaped by radiative feedback}",
      journal = {\nat},
         year = 2017,
        month = sep,
       volume = {549},
       number = {7673},
        pages = {488-491},
          doi = {10.1038/nature23906},
archivePrefix = {arXiv},
       eprint = {1709.09651},
 primaryClass = {astro-ph.HE},
       adsurl = {https://ui.adsabs.harvard.edu/abs/2017Natur.549..488R}
}

@ARTICLE{Ricci2022,
       author = {{Ricci}, C. and {Ananna}, T.~T. and {Temple}, M.~J. and {Urry}, C.~M. and {Koss}, M.~J. and {Trakhtenbrot}, B. and {Ueda}, Y. and {Stern}, D. and {Bauer}, F.~E. and {Treister}, E. and {Privon}, G.~C. and {Oh}, K. and {Paltani}, S. and {Stalevski}, M. and {Ho}, L.~C. and {Fabian}, A.~C. and {Mushotzky}, R. and {Chang}, C.~S. and {Ricci}, F. and {Kakkad}, D. and {Sartori}, L. and {Baer}, R. and {Caglar}, T. and {Powell}, M. and {Harrison}, F.},
        title = "{BASS XXXVII: The Role of Radiative Feedback in the Growth and Obscuration Properties of Nearby Supermassive Black Holes}",
      journal = {\apj},
         year = 2022,
        month = oct,
       volume = {938},
       number = {1},
          eid = {67},
        pages = {67},
          doi = {10.3847/1538-4357/ac8e67},
archivePrefix = {arXiv},
       eprint = {2209.00014},
 primaryClass = {astro-ph.GA},
       adsurl = {https://ui.adsabs.harvard.edu/abs/2022ApJ...938...67R}
}

@ARTICLE{Raban2009,
       author = {{Raban}, David and {Jaffe}, Walter and {R{\"o}ttgering}, Huub and {Meisenheimer}, Klaus and {Tristram}, Konrad R.~W.},
        title = "{Resolving the obscuring torus in NGC 1068 with the power of infrared interferometry: revealing the inner funnel of dust}",
      journal = {\mnras},
         year = 2009,
        month = apr,
       volume = {394},
       number = {3},
        pages = {1325-1337},
          doi = {10.1111/j.1365-2966.2009.14439.x},
archivePrefix = {arXiv},
       eprint = {0901.1306},
 primaryClass = {astro-ph.GA},
       adsurl = {https://ui.adsabs.harvard.edu/abs/2009MNRAS.394.1325R}
}

@ARTICLE{Honig2012,
       author = {{H{\"o}nig}, S.~F. and {Kishimoto}, M. and {Antonucci}, R. and {Marconi}, A. and {Prieto}, M.~A. and {Tristram}, K. and {Weigelt}, G.},
        title = "{Parsec-scale Dust Emission from the Polar Region in the Type 2 Nucleus of NGC 424}",
      journal = {\apj},
         year = 2012,
        month = aug,
       volume = {755},
       number = {2},
          eid = {149},
        pages = {149},
          doi = {10.1088/0004-637X/755/2/149},
archivePrefix = {arXiv},
       eprint = {1206.4307},
 primaryClass = {astro-ph.CO},
       adsurl = {https://ui.adsabs.harvard.edu/abs/2012ApJ...755..149H}
}

@ARTICLE{Honig2013,
       author = {{H{\"o}nig}, S.~F. and {Kishimoto}, M. and {Tristram}, K.~R.~W. and {Prieto}, M.~A. and {Gandhi}, P. and {Asmus}, D. and {Antonucci}, R. and {Burtscher}, L. and {Duschl}, W.~J. and {Weigelt}, G.},
        title = "{Dust in the Polar Region as a Major Contributor to the Infrared Emission of Active Galactic Nuclei}",
      journal = {\apj},
         year = 2013,
        month = jul,
       volume = {771},
       number = {2},
          eid = {87},
        pages = {87},
          doi = {10.1088/0004-637X/771/2/87},
archivePrefix = {arXiv},
       eprint = {1306.4312},
 primaryClass = {astro-ph.CO},
       adsurl = {https://ui.adsabs.harvard.edu/abs/2013ApJ...771...87H}
}

@ARTICLE{Lopez-Gonzaga2014,
       author = {{L{\'o}pez-Gonzaga}, N. and {Jaffe}, W. and {Burtscher}, L. and {Tristram}, K.~R.~W. and {Meisenheimer}, K.},
        title = "{Revealing the large nuclear dust structures in NGC 1068 with MIDI/VLTI}",
      journal = {\aap},
         year = 2014,
        month = may,
       volume = {565},
          eid = {A71},
        pages = {A71},
          doi = {10.1051/0004-6361/201323002},
archivePrefix = {arXiv},
       eprint = {1401.3248},
 primaryClass = {astro-ph.GA},
       adsurl = {https://ui.adsabs.harvard.edu/abs/2014A&A...565A..71L}
}

@ARTICLE{Lopez-Gonzaga2016,
       author = {{L{\'o}pez-Gonzaga}, N. and {Burtscher}, L. and {Tristram}, K.~R.~W. and {Meisenheimer}, K. and {Schartmann}, M.},
        title = "{Mid-infrared interferometry of 23 AGN tori: On the significance of polar-elongated emission}",
      journal = {\aap},
         year = 2016,
        month = jun,
       volume = {591},
          eid = {A47},
        pages = {A47},
          doi = {10.1051/0004-6361/201527590},
archivePrefix = {arXiv},
       eprint = {1602.05592},
 primaryClass = {astro-ph.GA},
       adsurl = {https://ui.adsabs.harvard.edu/abs/2016A&A...591A..47L}
}

@ARTICLE{Leftley2018,
       author = {{Leftley}, James H. and {Tristram}, Konrad R.~W. and {H{\"o}nig}, Sebastian F. and {Kishimoto}, Makoto and {Asmus}, Daniel and {Gandhi}, Poshak},
        title = "{New Evidence for the Dusty Wind Model: Polar Dust and a Hot Core in the Type-1 Seyfert ESO 323-G77}",
      journal = {\apj},
         year = 2018,
        month = jul,
       volume = {862},
       number = {1},
          eid = {17},
        pages = {17},
          doi = {10.3847/1538-4357/aac8e5},
archivePrefix = {arXiv},
       eprint = {1806.01863},
 primaryClass = {astro-ph.GA},
       adsurl = {https://ui.adsabs.harvard.edu/abs/2018ApJ...862...17L}
}

@ARTICLE{Hopkins2008,
       author = {{Hopkins}, Philip F. and {Hernquist}, Lars and {Cox}, Thomas J. and {Kere{\v{s}}}, Du{\v{s}}an},
        title = "{A Cosmological Framework for the Co-Evolution of Quasars, Supermassive Black Holes, and Elliptical Galaxies. I. Galaxy Mergers and Quasar Activity}",
      journal = {\apjs},
         year = 2008,
        month = apr,
       volume = {175},
       number = {2},
        pages = {356-389},
          doi = {10.1086/524362},
archivePrefix = {arXiv},
       eprint = {0706.1243},
 primaryClass = {astro-ph},
       adsurl = {https://ui.adsabs.harvard.edu/abs/2008ApJS..175..356H}
}

@ARTICLE{Hickox2018,
       author = {{Hickox}, Ryan C. and {Alexander}, David M.},
        title = "{Obscured Active Galactic Nuclei}",
      journal = {\araa},
         year = 2018,
        month = sep,
       volume = {56},
        pages = {625-671},
          doi = {10.1146/annurev-astro-081817-051803},
archivePrefix = {arXiv},
       eprint = {1806.04680},
 primaryClass = {astro-ph.GA},
       adsurl = {https://ui.adsabs.harvard.edu/abs/2018ARA&A..56..625H}
}

@ARTICLE{Haidar2024,
       author = {{Haidar}, Houda and {Rosario}, David J. and {Alonso-Herrero}, Almudena and {Pereira-Santaella}, Miguel and {Garc{\'\i}a-Bernete}, Ismael and {Campbell}, Stephanie and {H{\"o}nig}, Sebastian F. and {Ramos Almeida}, Cristina and {Hicks}, Erin and {Delaney}, Daniel and {Davies}, Richard and {Ricci}, Claudio and {Harrison}, Chris M. and {Leist}, Mason and {Lopez-Rodriguez}, Enrique and {Garcia-Burillo}, Santiago and {Zhang}, Lulu and {Packham}, Chris and {Gandhi}, Poshak and {Audibert}, Anelise and {Bellocchi}, Enrica and {Boorman}, Peter and {Bunker}, Andrew and {Combes}, Fran{\c{c}}oise and {Diaz Santos}, Tanio and {Donnan}, Fergus R. and {Gonzalez Martin}, Omaira and {Hermosa Mu{\~n}oz}, Laura and {Charidis}, Matthaios and {Labiano}, Alvaro and {Levenson}, Nancy A. and {May}, Daniel and {Rigopoulou}, Dimitra and {Rodriguez Ardila}, Alberto and {Shimizu}, T. Taro and {Stalevski}, Marko and {Ward}, Martin},
        title = "{Dust beyond the torus: revealing the mid-infrared heart of local Seyfert ESO 428-G14 with JWST/MIRI}",
      journal = {\mnras},
         year = 2024,
        month = aug,
       volume = {532},
       number = {4},
        pages = {4645-4660},
          doi = {10.1093/mnras/stae1596},
archivePrefix = {arXiv},
       eprint = {2404.16100},
 primaryClass = {astro-ph.GA},
       adsurl = {https://ui.adsabs.harvard.edu/abs/2024MNRAS.532.4645H}
}

@ARTICLE{Kormendy2013,
       author = {{Kormendy}, John and {Ho}, Luis C.},
        title = "{Coevolution (Or Not) of Supermassive Black Holes and Host Galaxies}",
      journal = {\araa},
         year = 2013,
        month = aug,
       volume = {51},
       number = {1},
        pages = {511-653},
          doi = {10.1146/annurev-astro-082708-101811},
archivePrefix = {arXiv},
       eprint = {1304.7762},
 primaryClass = {astro-ph.CO},
       adsurl = {https://ui.adsabs.harvard.edu/abs/2013ARA&A..51..511K}
}

@ARTICLE{Tristram2014,
       author = {{Tristram}, Konrad R.~W. and {Burtscher}, Leonard and {Jaffe}, Walter and {Meisenheimer}, Klaus and {H{\"o}nig}, Sebastian F. and {Kishimoto}, Makoto and {Schartmann}, Marc and {Weigelt}, Gerd},
        title = "{The dusty torus in the Circinus galaxy: a dense disk and the torus funnel}",
      journal = {\aap},
         year = 2014,
        month = mar,
       volume = {563},
          eid = {A82},
        pages = {A82},
          doi = {10.1051/0004-6361/201322698},
archivePrefix = {arXiv},
       eprint = {1312.4534},
 primaryClass = {astro-ph.GA},
       adsurl = {https://ui.adsabs.harvard.edu/abs/2014A&A...563A..82T}
}

@ARTICLE{Garcia-Bernete2022c,
       author = {{Garc{\'\i}a-Bernete}, I. and {Rigopoulou}, D. and {Alonso-Herrero}, A. and {Donnan}, F.~R. and {Roche}, P.~F. and {Pereira-Santaella}, M. and {Labiano}, A. and {Peralta de Arriba}, L. and {Izumi}, T. and {Ramos Almeida}, C. and {Shimizu}, T. and {H{\"o}nig}, S. and {Garc{\'\i}a-Burillo}, S. and {Rosario}, D.~J. and {Ward}, M.~J. and {Bellocchi}, E. and {Hicks}, E.~K.~S. and {Fuller}, L. and {Packham}, C.},
        title = "{A high angular resolution view of the PAH emission in Seyfert galaxies using JWST/MRS data}",
      journal = {\aap},
         year = 2022,
        month = oct,
       volume = {666},
          eid = {L5},
        pages = {L5},
          doi = {10.1051/0004-6361/202244806},
archivePrefix = {arXiv},
       eprint = {2208.11620},
 primaryClass = {astro-ph.GA},
       adsurl = {https://ui.adsabs.harvard.edu/abs/2022A&A...666L...5G}
}

@ARTICLE{Garcia-Bernete2022b,
       author = {{Garc{\'\i}a-Bernete}, I. and {Rigopoulou}, D. and {Alonso-Herrero}, A. and {Pereira-Santaella}, M. and {Roche}, P.~F. and {Kerkeni}, B.},
        title = "{Polycyclic aromatic hydrocarbons in Seyfert and star-forming galaxies}",
      journal = {\mnras},
         year = 2022,
        month = jan,
       volume = {509},
       number = {3},
        pages = {4256-4275},
          doi = {10.1093/mnras/stab3127},
archivePrefix = {arXiv},
       eprint = {2011.10882},
 primaryClass = {astro-ph.GA},
       adsurl = {https://ui.adsabs.harvard.edu/abs/2022MNRAS.509.4256G}
}

@ARTICLE{RamosAlmeida2019,
       author = {{Ramos Almeida}, C. and {Acosta-Pulido}, J.~A. and {Tadhunter}, C.~N. and {Gonz{\'a}lez-Fern{\'a}ndez}, C. and {Cicone}, C. and {Fern{\'a}ndez-Torreiro}, M.},
        title = "{A near-infrared study of the multiphase outflow in the type-2 quasar J1509+0434}",
      journal = {\mnras},
         year = 2019,
        month = jul,
       volume = {487},
       number = {1},
        pages = {L18-L23},
          doi = {10.1093/mnrasl/slz072},
archivePrefix = {arXiv},
       eprint = {1905.06288},
 primaryClass = {astro-ph.GA},
       adsurl = {https://ui.adsabs.harvard.edu/abs/2019MNRAS.487L..18R}
}

@ARTICLE{Mingozzi2019,
       author = {{Mingozzi}, M. and {Cresci}, G. and {Venturi}, G. and {Marconi}, A. and {Mannucci}, F. and {Perna}, M. and {Belfiore}, F. and {Carniani}, S. and {Balmaverde}, B. and {Brusa}, M. and {Cicone}, C. and {Feruglio}, C. and {Gallazzi}, A. and {Mainieri}, V. and {Maiolino}, R. and {Nagao}, T. and {Nardini}, E. and {Sani}, E. and {Tozzi}, P. and {Zibetti}, S.},
        title = "{The MAGNUM survey: different gas properties in the outflowing and disc components in nearby active galaxies with MUSE}",
      journal = {\aap},
         year = 2019,
        month = feb,
       volume = {622},
          eid = {A146},
        pages = {A146},
          doi = {10.1051/0004-6361/201834372},
archivePrefix = {arXiv},
       eprint = {1811.07935},
 primaryClass = {astro-ph.GA},
       adsurl = {https://ui.adsabs.harvard.edu/abs/2019A&A...622A.146M}
}

@ARTICLE{Fischer2018,
       author = {{Fischer}, Travis C. and {Kraemer}, S.~B. and {Schmitt}, H.~R. and {Longo Micchi}, L.~F. and {Crenshaw}, D.~M. and {Revalski}, M. and {Vestergaard}, M. and {Elvis}, M. and {Gaskell}, C.~M. and {Hamann}, F. and {Ho}, L.~C. and {Hutchings}, J. and {Mushotzky}, R. and {Netzer}, H. and {Storchi-Bergmann}, T. and {Straughn}, A. and {Turner}, T.~J. and {Ward}, M.~J.},
        title = "{Hubble Space Telescope Observations of Extended [O III]{\ensuremath{\lambda}} 5007 Emission in Nearby QSO2s: New Constraints on AGN Host Galaxy Interaction}",
      journal = {\apj},
         year = 2018,
        month = apr,
       volume = {856},
       number = {2},
          eid = {102},
        pages = {102},
          doi = {10.3847/1538-4357/aab03e},
archivePrefix = {arXiv},
       eprint = {1802.06184},
 primaryClass = {astro-ph.GA},
       adsurl = {https://ui.adsabs.harvard.edu/abs/2018ApJ...856..102F}
}

@ARTICLE{Ulivi2024,
       author = {{Ulivi}, L. and {Venturi}, G. and {Cresci}, G. and {Marconi}, A. and {Marconcini}, C. and {Amiri}, A. and {Belfiore}, F. and {Bertola}, E. and {Carniani}, S. and {D'Amato}, Q. and {Di Teodoro}, E. and {Ginolfi}, M. and {Girdhar}, A. and {Harrison}, C. and {Maiolino}, R. and {Mannucci}, F. and {Mingozzi}, M. and {Perna}, M. and {Scialpi}, M. and {Tomicic}, N. and {Tozzi}, G. and {Treister}, E.},
        title = "{Feedback and ionized gas outflows in four low-radio power AGN at z {\ensuremath{\sim}} 0.15}",
      journal = {\aap},
         year = 2024,
        month = may,
       volume = {685},
          eid = {A122},
        pages = {A122},
          doi = {10.1051/0004-6361/202347436},
archivePrefix = {arXiv},
       eprint = {2403.01258},
 primaryClass = {astro-ph.GA},
       adsurl = {https://ui.adsabs.harvard.edu/abs/2024A&A...685A.122U}
}

@ARTICLE{Zanchettin2025,
       author = {{Zanchettin}, M.~V. and {Ramos Almeida}, C. and {Audibert}, A. and {Acosta-Pulido}, J.~A. and {Cezar}, P.~H. and {Hicks}, E. and {Lapi}, A. and {Mullaney}, J.},
        title = "{Unveiling the warm molecular outflow component of type-2 quasars with SINFONI}",
      journal = {\aap},
         year = 2025,
        month = mar,
       volume = {695},
          eid = {A185},
        pages = {A185},
          doi = {10.1051/0004-6361/202453224},
archivePrefix = {arXiv},
       eprint = {2502.12800},
 primaryClass = {astro-ph.GA},
       adsurl = {https://ui.adsabs.harvard.edu/abs/2025A&A...695A.185Z}
}

@ARTICLE{Harrison2015,
       author = {{Harrison}, C.~M. and {Thomson}, A.~P. and {Alexander}, D.~M. and {Bauer}, F.~E. and {Edge}, A.~C. and {Hogan}, M.~T. and {Mullaney}, J.~R. and {Swinbank}, A.~M.},
        title = "{Storm in a ``Teacup'': A Radio-quiet Quasar with {\ensuremath{\approx}}10 kpc Radio-emitting Bubbles and Extreme Gas Kinematics}",
      journal = {\apj},
         year = 2015,
        month = feb,
       volume = {800},
       number = {1},
          eid = {45},
        pages = {45},
          doi = {10.1088/0004-637X/800/1/45},
archivePrefix = {arXiv},
       eprint = {1410.4198},
 primaryClass = {astro-ph.GA},
       adsurl = {https://ui.adsabs.harvard.edu/abs/2015ApJ...800...45H}
}

@ARTICLE{Venturi2023,
       author = {{Venturi}, G. and {Treister}, E. and {Finlez}, C. and {D'Ago}, G. and {Bauer}, F. and {Harrison}, C.~M. and {Ramos Almeida}, C. and {Revalski}, M. and {Ricci}, F. and {Sartori}, L.~F. and {Girdhar}, A. and {Keel}, W.~C. and {Tub{\'\i}n}, D.},
        title = "{Complex AGN feedback in the Teacup galaxy. A powerful ionised galactic outflow, jet-ISM interaction, and evidence for AGN-triggered star formation in a giant bubble}",
      journal = {\aap},
         year = 2023,
        month = oct,
       volume = {678},
          eid = {A127},
        pages = {A127},
          doi = {10.1051/0004-6361/202347375},
archivePrefix = {arXiv},
       eprint = {2309.02498},
 primaryClass = {astro-ph.GA},
       adsurl = {https://ui.adsabs.harvard.edu/abs/2023A&A...678A.127V}
}

@ARTICLE{Veenema2026,
       author = {{Veenema}, Oscar and {Thatte}, Niranjan and {Rigopoulou}, Dimitra and {Garc{\'\i}a-Bernete}, Ismael and {Alonso-Herrero}, Almudena and {Pereira-Santaella}, Miguel and {Audibert}, Anelise and {Bellocchi}, Enrica and {Bunker}, Andrew J. and {Campbell}, Steph and {Combes}, Francoise and {Davies}, Richard I. and {Donnan}, Fergus R. and {Garc{\'\i}a-Burillo}, Santiago and {Gonzalez Martin}, Omaira and {Hermosa Mu{\~n}oz}, Laura and {Hicks}, Erin K.~S. and {Hoenig}, Sebastian F. and {Labiano}, Alvaro and {Levenson}, Nancy A. and {Packham}, Chris and {Ramos Almeida}, Cristina and {Ricci}, Claudio and {Riffel}, Rogemar A. and {Rosario}, David and {Shimizu}, Taro and {Zhang}, Lulu},
        title = "{Decoupling the AGN outflow and star-forming disc kinematics in the nuclear region of NGC 7582 with JWST NIRSpec and MIRI/MRS}",
      journal = {\mnras},
         year = 2026,
        month = jun,
       volume = {548},
       number = {4},
          eid = {stag785},
        pages = {stag785},
          doi = {10.1093/mnras/stag785},
archivePrefix = {arXiv},
       eprint = {2604.24892},
 primaryClass = {astro-ph.GA},
       adsurl = {https://ui.adsabs.harvard.edu/abs/2026MNRAS.548ag785V}
}

@ARTICLE{Diamond-Stanic2010,
       author = {{Diamond-Stanic}, Aleksandar M. and {Rieke}, George H.},
        title = "{The Effect of Active Galactic Nuclei on the Mid-infrared Aromatic Features}",
      journal = {\apj},
         year = 2010,
        month = nov,
       volume = {724},
       number = {1},
        pages = {140-153},
          doi = {10.1088/0004-637X/724/1/140},
archivePrefix = {arXiv},
       eprint = {1009.2752},
 primaryClass = {astro-ph.CO},
       adsurl = {https://ui.adsabs.harvard.edu/abs/2010ApJ...724..140D}
}

@ARTICLE{ODowd2009,
       author = {{O'Dowd}, Matthew J. and {Schiminovich}, David and {Johnson}, Benjamin D. and {Treyer}, Marie A. and {Martin}, Christopher D. and {Wyder}, Ted K. and {Charlot}, S. and {Heckman}, Timothy M. and {Martins}, Lucimara P. and {Seibert}, Mark and {van der Hulst}, J.~M.},
        title = "{Polycyclic Aromatic Hydrocarbons in Galaxies at z \raisebox{-0.5ex}\textasciitilde 0.1: The Effect of Star Formation and Active Galactic Nuclei}",
      journal = {\apj},
         year = 2009,
        month = nov,
       volume = {705},
       number = {1},
        pages = {885-898},
          doi = {10.1088/0004-637X/705/1/885},
archivePrefix = {arXiv},
       eprint = {0909.2279},
 primaryClass = {astro-ph.CO},
       adsurl = {https://ui.adsabs.harvard.edu/abs/2009ApJ...705..885O}
}

@ARTICLE{Bianchin2026,
       author = {{Bianchin}, M. and {Ramos Almeida}, C. and {Gonz{\'a}lez-Mart{\'\i}n}, O. and {Zanchettin}, M.~V. and {Carneiro}, M. and {Pereira-Santaella}, M. and {Tadhunter}, C. and {Speranza}, G. and {Garc{\'\i}a-Bernete}, I. and {Audibert}, A. and {Alonso-Herrero}, A. and {Rigopoulou}, D. and {Labiano}, A. and {Acosta-Pulido}, J.~A. and {Garc{\'\i}a-Burillo}, S.},
        title = "{Extended coronal line emission and new clues to a possible dual AGN in the merger J1356+1026}",
      journal = {arXiv e-prints},
         year = 2026,
        month = apr,
          eid = {arXiv:2604.08239},
        pages = {arXiv:2604.08239},
          doi = {10.48550/arXiv.2604.08239},
archivePrefix = {arXiv},
       eprint = {2604.08239},
 primaryClass = {astro-ph.GA},
       adsurl = {https://ui.adsabs.harvard.edu/abs/2026arXiv260408239B}
}

@ARTICLE{Sanders1988,
       author = {{Sanders}, D.~B. and {Soifer}, B.~T. and {Elias}, J.~H. and {Neugebauer}, G. and {Matthews}, K.},
        title = "{Warm Ultraluminous Galaxies in the IRAS Survey: The Transition from Galaxy to Quasar?}",
      journal = {\apjl},
         year = 1988,
        month = may,
       volume = {328},
        pages = {L35},
          doi = {10.1086/185155},
       adsurl = {https://ui.adsabs.harvard.edu/abs/1988ApJ...328L..35S}
}
\bibliographystyle{aa}

\begin{appendix}
\nolinenumbers
\section{Comparison to line fitting}
\label{sec:LineFits}
To verify that the PCA technique is accurate at producing velocity maps, we compare it to a more traditional method of fitting a Gaussian line profile to every spaxel for the emission lines where such traditional methods are viable. For each spaxel, we allow the amplitude, central wavelength and standard deviation of the Gaussian to vary as free parameters, where we use the central wavelength to infer the velocity. 

\begin{figure}
        \includegraphics[width=\columnwidth]{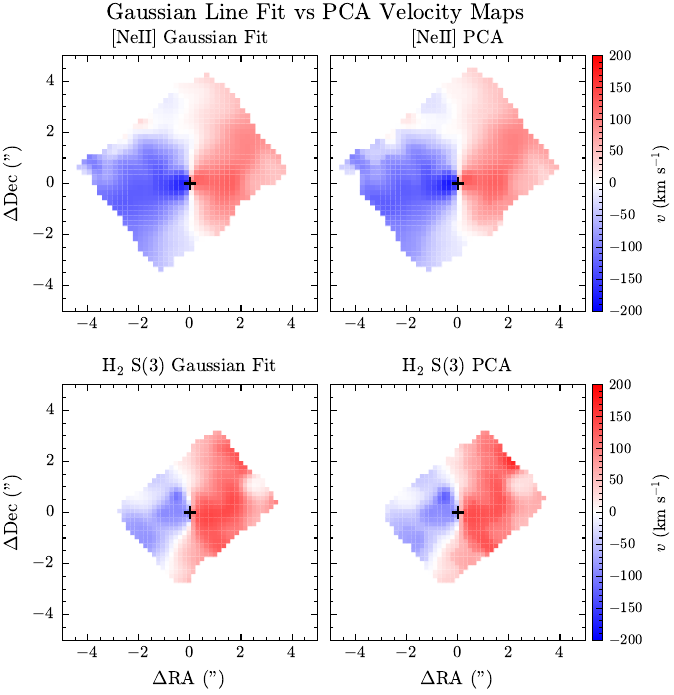}
    \caption{Comparison between the PCA technique and a more traditional method of producing a velocity map by fitting a Gaussian to the spectrum of each spaxel. We show this comparison for the [NeII] and H$_2$ S(3) line for J1509.}
    \label{fig:LinePlot} 
\end{figure}

We show a comparison in Fig. \ref{fig:LinePlot} for J1509 where we show [NeII] and the H$_2$ S(3) line. We find that the velocities are almost identical with the PCA doing a good job at accurately inferring the velocities.

\section{Continuum subtraction}
\label{sec:ContSub}
To investigate the impact that our choice of continuum subtraction has on the PCA decomposition and the subsequent velocities we measure, we test different continuum subtraction methods. We use the 11.3 $\mu$m PAH feature in J1509 for this testing.

\begin{figure}
        \includegraphics[width=\columnwidth]{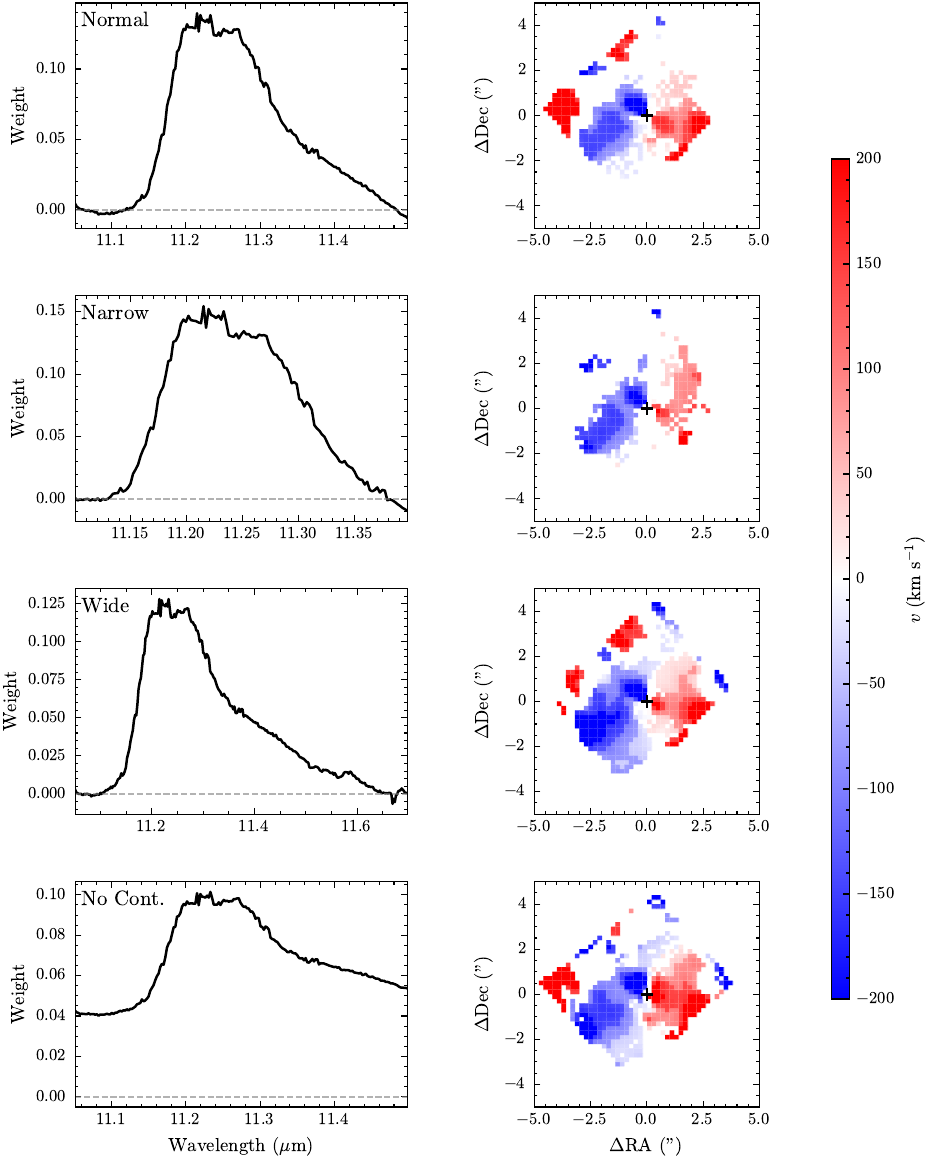}
    \caption{Tests of different continuum subtractions for the 11.3 $\mu$m PAH feature in J1509. \textit{Left panels}: Eigenspectra of the first principal component. \textit{Right panels}: Inferred velocity map from the PCA decomposition. The top panels show the continuum subtraction used in this work, while the following panels show a narrower wavelength range, a wider wavelength range, and no continuum subtraction.}
    \label{fig:ContTest} 
\end{figure}

In Fig. \ref{fig:ContTest}, we show different continuum subtractions and the resulting velocity maps. In particular we try a narrower wavelength range where we exclude more of the wings of the feature. We also try a wider wavelength range, where we include more of the wing on the longer wavelength side. Finally we try no continuum subtraction at all. 

We find that in all cases the velocity map is broadly consistent which suggests that the continuum subtraction method does not have a large impact on the resultant velocity map. Even when the wavelength range is very narrow and the continuum subtraction removes much of the PAH emission from the wings of the feature, the core is sufficient to measure kinematics from. Therefore the continuum subtraction we have used in this work is sufficient to measure the kinematics of the 11.3 $\mu$m PAh feature.

We also test no continuum subtraction, where the kinematics are surprisingly well constrained. This is likely because the continuum is weak compared to the PAH feature, and the changes in the profile due to a Doppler shift still dominates the higher order principal components.

\end{appendix}
\end{document}